\documentclass[fleqn,usenatbib,useAMS]{mnras}

\usepackage{graphicx}	
\usepackage{amsmath}	
\usepackage{amssymb}	
\usepackage{multicol}        
\usepackage{bm}		
\usepackage{pdflscape}	
\usepackage{multirow}
\usepackage{comment}
\usepackage{array}
\includecomment{comment} 

\newcommand\T{\rule{0pt}{2.6ex}}       
\newcommand\B{\rule[-1.2ex]{0pt}{0pt}} 

\newcommand{\rg}{\,r_g}

\usepackage[T1]{fontenc}
\usepackage{ae,aecompl}

\usepackage{caption}
\usepackage{subcaption}
\usepackage{float}

\title[Variability in GRMHD simulations]{Looking for the underlying cause of black hole X-ray variability in GRMHD simulations}

\author[Bollimpalli et al]{D. A. Bollimpalli$^{1,2}$\thanks{Contact e-mail: \href{mailto:deepika@camk.edu.pl}{deepika@camk.edu.pl}},
R. Mahmoud$^{3}$, 
C. Done$^{3}$, 
P. C. Fragile$^{2,4}$, 
\newauthor
W. Klu\'zniak$^{1}$, R. Narayan$^{5}$, C. J. White$^{4}$
\\
$^{1}$Nicolaus Copernicus Astronomical Center, Polish Academy of Sciences, ul. Bartycka 18, PL 00-716 Warsaw, Poland\\
$^{2}$Department of Physics and Astronomy, College of Charleston, Charleston, SC 29424, USA\\
$^{3}$Department of Physics, University of Durham, South Road, Durham DH1 3LE, UK\\
$^{4}$Kavli Institute for Theoretical Physics, Kohn Hall, University of California, Santa Barbara, CA 93107, USA\\
$^{5}$ Harvard-Smithsonian Center for Astrophysics, 60 Garden Street, Cambridge, MA 02138, USA}
\date{}

\pubyear{2019}
\begin{document}
\label{firstpage}
\pagerange{\pageref{firstpage}--\pageref{lastpage}}
\maketitle

\begin{abstract}
Long-term observations have shown that black hole X-ray binaries exhibit strong, aperiodic variability on time-scales of a few milliseconds to seconds. The observed light curves display various characteristic features like a log-normal distribution of flux and a linear rms-flux relation, which indicate that the underlying variability process is stochastic in nature. It is also thought to be intrinsic to accretion. This variability has been modelled as inward propagating fluctuations of mass accretion rate, although the physical process driving the fluctuations remains puzzling. In this work, we analyse five exceptionally long duration general relativistic magnetohydrodynamic (GRMHD) simulations of optically thin, geometrically thick, black hole accretion flows to look for hints of propagating fluctuations in the simulation data. We find that the accretion profiles from these simulations do show evidence for inward propagating fluctuations below the viscous frequency by featuring strong radial coherence and positive time lags when comparing smaller to larger radii, although these time lags are generally shorter than the viscous time-scale and frequency independent. Our simulations also support the notion that the fluctuations in $\dot{M}$ build up in a multiplicative manner, as the simulations exhibit linear rms-mass flux relations, as well as log-normal distributions of their mass fluxes. 
When combining the mass fluxes from the simulations with an assumed emissivity profile, we additionally find broad agreement with observed power spectra and time lags, including a recovery of the frequency dependency of the time lags.
\end{abstract}



\section{Introduction}
\label{sec:intro}
A large fraction of the galactic black hole X-ray binaries (BHXRBs, e.g. GX 339--4, XTE J1550--64, and GRO J1655--40) are generically variable, both spectrally and temporally, as may be associated with changes to the geometry and nature of accretion flows. They occasionally undergo outbursts during which the luminosity varies by several orders of magnitude, and the X-ray spectral states evolve  -- a phenomena termed ``state transitions.'' The two main spectral states seen in these systems are a high-luminosity, ``soft" state, dominated by thermal emission from the accretion disc; and a variable-luminosity, ``hard" state, dominated by a non-thermal Comptonised X-ray spectrum, with a weak or absent thermal component \citep[see, e.g., the review by][]{DGK07}. 

The X-ray light curves of BHXRBs, particularly in the hard state, are dominated by strong, aperiodic variability (flicker-type noise) on time-scales of milliseconds to seconds \citep{vanderklis95}. This manifests as a continuous band of power present over a wide range of frequencies in the power spectral densities (PSD), which also change during the state transitions. The other important variability features reported in these systems are a linear rms-flux relation and log-normal distribution of flux, indicating that the underlying stochastic process could be non-linear and multiplicative in nature \citep{UM2001, UMV2005}. Further hints on the origin of variability may be derived from the frequency-dependent time lags and strong coherence observed between different energy bands \citep{MK89, NVBW99, MPUPM04}. Similar variability features are observed to occur in active galactic nuclei (AGN) on time-scales of hours to months \citep{Mchardy88, G04}. 

In addition to the black hole systems, broad-band variability is also observed in other accreting systems: neutron star systems (in X-rays) \citep{WV99}, accreting white dwarfs (in optical/UV band) \citep{Scaringi13, VSK15} and young stellar objects (in IR/optical band) \citep{Scaringi15}. The remarkable similarity in the variability across different systems suggests that the underlying physical mechanism at work may be the same in all these systems.

\citet{L97} suggested a model where fluctuations of viscosity generated in the disc at different radii on viscous time-scales cause mass accretion rate ($\dot{M}$) fluctuations that propagate inwards by modulating subsequent fluctuations driven at smaller radii. Since the fluctuations couple together, this model readily explains multiple observed features: a) the accretion rate, and thereby the emission flux from the inner-regions, carries the imprints of fluctuations generated over a wide range of radii associated with a wide range of variability time-scales, thus producing a broad PSD; and b) the high-frequency fluctuations presumed to be driven at smaller radii are modulated by the longer time-scale fluctuations coming from larger radii, thus giving rise to the rms-flux relation over all time-scales. Furthermore, this coupling through a multiplicative combination of fluctuations naturally leads to a log-normal distribution. The observed energy-dependent PSDs, where the higher photon-energy bands show relatively more high-frequency power than lower photon-energy bands, can be explained by this model if the hard X-rays are produced from a concentrated region close to the central object and the soft X-rays come from an extended region further away. This basic picture also explains the strong coherence and time lags that appear between the different X-ray energy bands -- as the fluctuations propagate inwards, they first appear in the softer energy bands and later in the harder energy bands \citep{KCG01, MD18}. Additionally, outward propagating fluctuations from inner radii could potentially cause negative lags at higher frequencies, whether they are carried by waves \citep{Mushtukov18} or matter.\footnote{The velocity field in three-dimensional discs is nontrivial and even thin alpha-discs are known to exhibit backflows \citep{Urpin1984,KK2000,RG02,PR2017}} Since the accretion process carrying the fluctuations inwards is diffusive in nature, any fluctuations generated on time-scales shorter than the diffusion time-scales will be damped before they can reach the inner radii. Thus, to explain the observed high amplitude, high frequency variability power in the PSDs of BHXRBs may require the relatively short characteristic time-scales associated with geometrically thick, optically thin accretion flows \citep{CGR01,AU06, ID2011, MD18}.

Our general understanding of accretion discs is largely based on the concepts of \citet{SS73}, which describes geometrically thin and optically thick discs, emitting thermal blackbody-like radiation. However, that model was soon found to be incapable of producing the observed hard X-rays ($\sim$100 keV) from Cygnus X-1 \citep{LS75}. Various models were invoked to explain the observed spectral properties through hot, optically thin and geometrically thick flows at low luminosities, in the inner disc regions around the black hole \citep{TP75, SLE76}, which were later modified to include advection to ensure thermal stability and became what is known as ADAFs \citep[advection dominated accretion flows,][]{ichimaru77, NY95b}. In these flows, the energy released through viscous dissipation is stored in the accreted matter and advected radially into the black hole \citep{NY94}. Thus these flows are hot and radiatively inefficient, therefore sometimes are referred to as RIAFs (radiatively inefficient accretion flows). The typical temperatures in ADAFs are close to virial and thus have the potential to explain the observed hard spectra through inverse Compton scattering by hot electrons. These flows are, however, susceptible to the convective instability, which can play a vital role in launching outflows \citep{NY95a}. Further studies have developed analytical solutions that treat convection as a dominant process in transporting angular momentum \citep{NIA2000, QG2000}; such solutions are sometimes referred to as convection dominated accretion flows (CDAFs). Disc winds could also carry away the angular momentum from the accretion flow, in which case the net advected energy onto the black hole is reduced \citep{BB99}. It is now widely accepted that ADAFs can consistently account for the observations of a number of low-luminosity state BHXRBs and AGN, including Sgr A* \citep{RBBP82, NMGPG98}. 

All these ADAF models \citep[see e.g.,][]{YN2014} are self-similar analytic solutions derived primarily by assuming a constant $\alpha$-viscosity parameter. However, real accretion discs are magnetized and thought to be subject to the magnetorotational instability  \citep[MRI,][]{BH91}, which acts as a natural source of turbulence, mediating the outward transport of angular momentum, thus allowing for accretion.  

Many numerical simulations of ADAFs have been performed so far in the interest of understanding the details of their flow dynamics and their observational relevance, yet the presence of aperiodic variability in these simulations remains relatively unexplored.
Only recent MHD simulations of geometrically thin discs\footnote{These discs are thought to be more relevant to X-ray binaries in the high/soft state, white dwarfs systems and young stellar objects.} have been examined in such a way \cite{sarah2014,HR16}. 

In our work, we are motivated to search for aperiodic variability in ADAFs using GRMHD simulations for at least two main reasons: a) the association of broad-band variability with hard X-ray flux in BHXRBs suggests they come from ADAF discs; and b) the characteristic time-scales of standard, thin discs cannot explain the high amplitude, high frequency variability power in the PSDs of BHXRBs \citep{CGR01}. The GRMHD simulations analysed in this work are notable for their extremely long time duration, which makes it possible to probe a wider frequency range of broad-band variability. However, a drawback of the fast time-scales associated with ADAFs is that simulations of them can be particularly sensitive to the initial conditions used \citep{WQG2020}; for this reason, we analyse simulations with different initial conditions, which can help discern the robustness of the propagating fluctuations in these flows.  

Our paper is organized as follows. We provide brief descriptions of the simulations used in this analysis in Section~\ref{sec:data}. Since we use the mass accretion rate ($\dot{M}$) as a proxy for the luminosity, we describe our analysis of $\dot{M}$ in Section~\ref{sec:analysis}. In Section~\ref{sec:results}, we present our results in the form of power spectra, coherence plots, time lags, rms-$\dot{M}$ relations and $\dot{M}$ distributions. In Section~\ref{sec:obs}, we compare our results with observations and conclude in Section~\ref{sec:conclusions}.

\section{Description of simulations }
\label{sec:data}
We analyse the accretion rate data from five different simulations of ADAFs, which were initiated using different initial conditions and were performed using different GRMHD codes. While we performed simulation D in the interest of this work, simulations A, B and C were taken from \citet{WQG2020}, while simulation R comes from \citet{NSPK2012}. In this section, we briefly describe the numerical setups of each simulation and discuss their key differences. For more details of simulations A, B, C, and R, we refer the reader to the original papers.

First, we describe the similarities between the simulations before we delve into the differences. All five simulations are non-radiative, which allows us to scale our results to any required black hole mass. However, this implies that we do not have a direct estimation of luminosity from the simulations, limiting us to use the mass accretion rate as a proxy for luminosity/flux. All the simulations are initiated from a rotating, hydrostatic equilibrium 
torus of matter governed by gravity, pressure and centrifugal forces. 
The mass of the torus is assumed negligible compared to the mass of the 
black hole so that the gravity is fixed by the black hole spacetime. Each 
torus is threaded with weak magnetic fields that are susceptible to the MRI \citep{BH91}. Once the simulations start, the MRI grows and produces turbulence, which is responsible for transporting angular momentum outwards, allowing matter to accrete inwards. Thus the inner region of the torus turns into an accretion flow while the outer region acts as a matter reservoir, feeding the accretion flow 
throughout the simulation. Below, we provide details of the equilibrium 
torus solutions and the magnetic field configurations used for the initial setup of each simulation. Key simulation parameters are summarised in Table~\ref{tab:sim_info}.

\subsection*{Simulations A, B and C}
These simulations are performed using the GRMHD code, Athena++ \citep{WSG16, STW2020}. They are initialised using the torus solution of \citet{FM76}, in which $u_{\phi}u^t$ (in Boyer Lindquist coordinates) is constant. The torus has a pressure maximum located at $r_{\rm max} = 52\rg$, where $\rg=GM/c^2$ is the gravitational radius. The peak density at  $r_{\rm max}$ is normalised to $1$. Simulation~A is run for the longest time period, up to $t = 4.4\times 10^{5}\,GM/c^3$, while the variants of this simulation -- B and C -- are run up to $t = 2.2\times 10^{5}\,GM/c^3$. For A and B, the inner edge of the torus is set at $r_{\rm in} = 25\rg$, while for C, $r_{\rm in} = 25.1\rg$. A polytropic equation of state is used in each case, with an adiabatic index of $\Gamma =4/3$ for simulations A and B and $\Gamma = 5/3$ for simulation C. The initial magnetic field is chosen to be purely poloidal with the number of alternating polarity loops set along the radial and polar directions, $N_{\rm r}$ and $N_{\theta}$, different for each simulation. For simulations A and C, $N_{\rm r} = 6$ and $N_{\theta}=4$, while for simulation B, $N_{\rm r} = 6$ and $N_{\theta}=1$. Using the alternating poloidal loops prevents the accumulation of large net magnetic flux in the accretion flow. 

These simulations are evolved in the spacetime of a non-spinning black hole ($a_*=Jc/GM^2=0$) using spherical Kerr-Schild coordinates. Grid cells are logarithmically spaced in the radial direction extending from $r = 1.7\rg$ to $10^4\rg$ and compressed towards the equator in the polar direction to increase the resolution close to the symmetry plane of the disk. The base grid, covering the entire sphere, consists of $120\times20\times20$ cells in the radial, polar and azimuthal directions with an additional one level refinement introduced in the region $\theta \in (\pi/5,4\pi/5)$ and another level of refinement on top of that in the region $\theta \in (3\pi/10,7\pi/10)$.

The key finding of these simulations is that the resulting flow 
structure of ADAFs is sensitive to the initial conditions. The 
larger magnetic field loops in simulation B build up a coherent 
vertical flux that drives polar outflows. For the same reason, 
simulation~B attains a ``semi-MAD'' (magnetically-arrested disc) state with the magnetic flux on the horizon reaching a limiting value for a brief period. Simulations A and C, on the other hand, exhibit polar inflows and 
equatorial backflows, which cannot be explained by 
convective stability/meridional circulation. Just the difference in $\Gamma$ causes a much more spherical distribution of mass in 
simulation~C compared to simulations~A and B which present a more 
standard disc picture, with high density near the equatorial region 
and low density close to the polar region. Further details of these 
simulations are provided in \citet{WQG2020}.

\subsection*{Simulation R}
This simulation is performed using the 3D GRMHD code, HARM \citep{GMT03, Mckinney06, MB09}. The simulation is set up using a torus solution similar to the Polish doughnut \citep{PKN13}, for which $r_{\rm in} =10\rg$ and the outer radius is $1000\rg$. The pressure maximum is located around $19.21\rg$. The angular momentum in the torus (-$u_{\phi}/u_{t}$) is constant within $42\rg$; beyond this radius, it is set to 71 percent of the Keplerian value. A polytropic equation of state is used with $\Gamma = 5/3$. The initial magnetic field is purely poloidal with eight centres of poloidal loops with alternating polarity set along the radial direction ($N_{\rm r} = 8$, $N_{\theta}=1$). 

This simulation is also evolved in the spacetime of a non-spinning black hole ($a_*=0$) using spherical Kerr-Schild coordinates. Grid cells are logarithmically spaced at smaller radii and hyper-logarithmically at the larger radii. Grid cells along the polar direction are non-uniformly spaced so as to increase the resolution close to both the equatorial region and the poles. This simulation uses a grid of $256\times128\times64$ cells in the radial, polar and azimuthal directions without any additional refinement. 

Simulation~R never reaches a MAD state even though the initial magnetic field configuration is similar to simulation~B. This simulation is convectively stable. Further details of this simulation are provided in \citet{NSPK2012}.

\subsection*{Simulation D}
This simulation is performed using the 3D GRMHD code, Cosmos++ \citep{AFS05, FGMRA12, FOA14}. The simulation is set up following the torus solution given by \citet{Chakrabarti85}, for which we set $r_{\rm in} =22\rg$ and $r_{\rm max}=40\rg$. The angular momentum inside the torus follows a power-law distribution in radius. The torus is threaded with a single poloidal magnetic field loop. The simulation is evolved in the spacetime of a spinning black hole with $a_*=0.5$ using spherical Kerr-Schild coordinates. A polytropic equation of state is used with $\Gamma = 4/3$. This simulation uses a grid of $192\times128\times32$ cells in the radial, polar and azimuthal directions without any additional refinement. For more details on this simulation, see Appendix \ref{appendix:a}.

\begin{table}
\begin{tabular}{m{0.25cm}m{0.25cm}m{0.35cm}m{0.35cm}m{0.35cm}m{2cm}m{1cm}m{1cm}}
\hline
Sim. & $a_*$& $r_{\rm max}$ & $r_{\rm in}$ & $\Gamma$ & B field & run & dump\\ 
& & & & & &length  &interval\\
 & & [$r_g$] & [$r_g$]  & &[number of loops]&[$r_g/c$] & [$r_g/c$]\T\B\\  \hline
A & 0 & 52 & 25.0& 4/3& $N_r = 6, N_{\theta}=4$ &440,000 & 100 (1)$^{\dagger}$ \T\B\\ \hline
B & 0 & 52 &25.0& 4/3&$N_r = 6, N_{\theta}=1$ &220,000 & 100 \T\B\\ \hline
C & 0 & 52 &25.1& 5/3&$N_r = 6, N_{\theta}=4$ &220,000 & 100 \T\B\\ \hline
R & 0 & 19.2 & 10.0& 5/3&$N_r = 8, N_{\theta}=1$ &200,000 & 10 \T\B\\  \hline
D & 0.5 & 40 &22.0& 4/3&$N_r = 1, N_{\theta}=1$&63,000 & 100 \T\B\\ \hline
\end{tabular}
\caption{Parameters of the simulations considered for analysis in this paper.\\$\dagger$ For this particular simulation, we also have the high cadence data with dumping interval equal to $1\,GM/c^3$ for variables at a few predetermined radii.}
\label{tab:sim_info}
\end{table}
\begin{figure*}
\centering
\includegraphics[width=\textwidth]{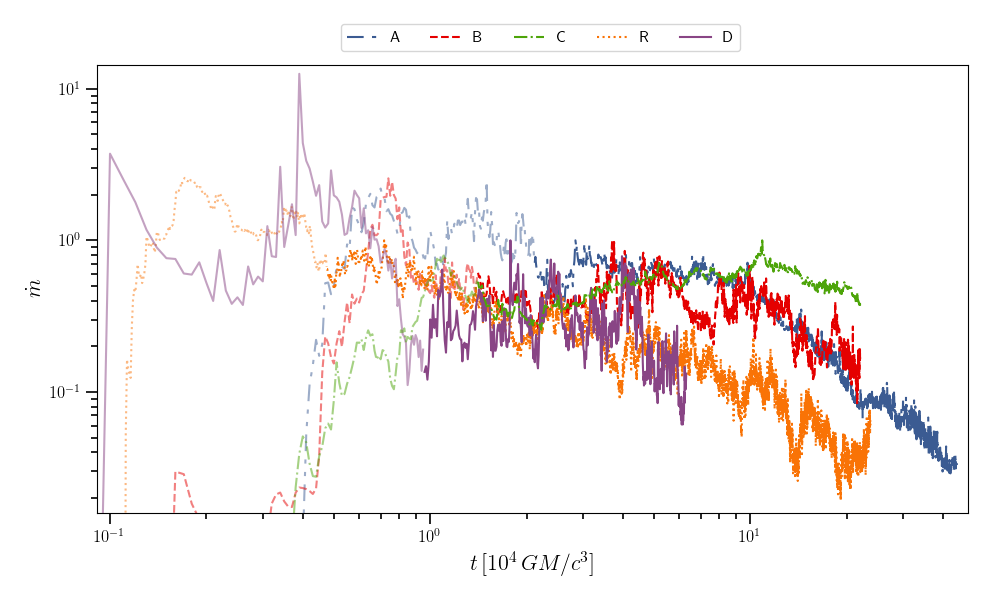}
\caption{Profiles of the normalized accretion rate, $\dot{m}$, through the black hole horizon for all five simulations. Discarded data for simulations B, C and D (first $6~P_{\rm orb}$) and simulations A and R (first $9~P_{\rm orb}$) are shown in lighter shades.}
\label{fig:mdot_rin}
\end{figure*}

\section{Accretion rate}
\label{sec:analysis}
The mass accretion rate, $\dot{M}$, at a given radius and time is computed by integrating the mass flux over a spherical shell: 
\begin{equation}
    \dot{M}(r,t) = -\iint\rho u^{r} \sqrt{-g}{\rm d}\theta {\rm d}\phi~,
\end{equation}
where $\rho$ is the rest mass density, $u^r$ is radial four-velocity, and the negative sign is included to make $\dot{M}$ positive when matter is flowing inwards towards the black hole. We calculate this expression in spherical Kerr-Schild coordinates. In our analysis, we discard the initial six orbital periods ($P_{\rm orb}$) of simulation data for simulations B, C and D and nine orbital periods for simulations A and R, where $P_{\rm orb}$ is measured at $r_{\rm max}$. $\dot{M}$ for each simulation is then re-normalised by its maximum value outside the discarded initial transient phase. In the rest of the paper we report the normalised accretion rate, $\dot{m} = \dot{M}/{\rm max}(\dot{M})|_{r_{\rm H}}$, where $r_{\rm H}$ is the horizon radius. 

Fig.~\ref{fig:mdot_rin} shows the evolution of the accretion rate onto the black hole for all five simulations. The discarded data, corresponding to the initial transient phase of the simulations, is shown in lighter shade. Simulation~C remains at a comparatively high accretion rates throughout, while simulations A and B show secular declines beyond $\approx 100\,000\,GM/c^3$, with the decrease in $\dot{m}$ nearly two orders of magnitude for simulation A. Simulation~R, as well, shows a significant decline in $\dot{m}$, while no such decline is seen in simulation~D, though this simulation is run for a shorter time period. This secular decline behaviour of $\dot{m}$ can be attributed to two principal causes: a) The disc loses its mass through the horizon at a rate faster than it is supplied by the surrounding torus  \citep[see e.g., fig 6 in][]{NSPK2012}; and b) since the torus mass is not replenished in any of the simulations, $\dot{m}$ must inevitably begin to decline at some point as the mass reservoir becomes drained. 

\begin{figure*}
\centering
\includegraphics[width=\textwidth]{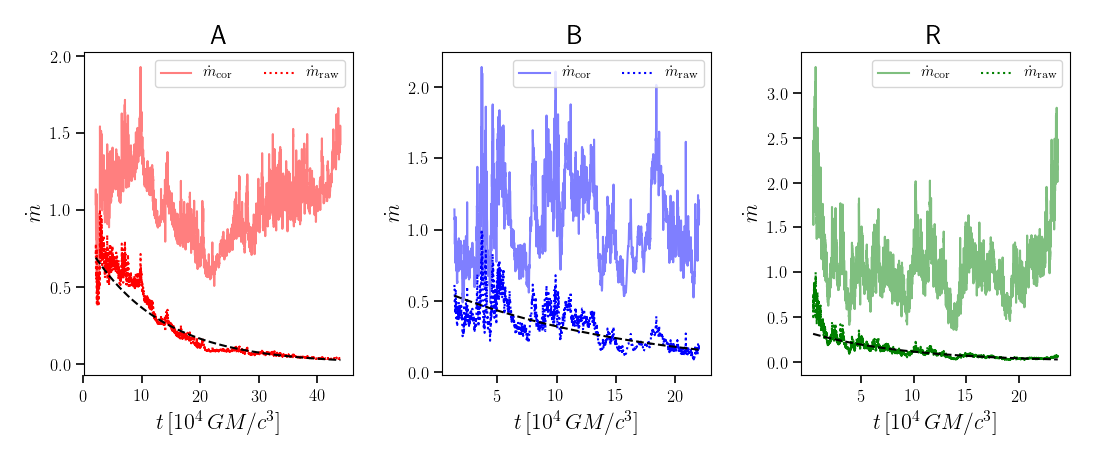}
\caption{Profiles of the mass accretion rates at $r_\mathrm{H}$ for simulations A (left), B (middle), and R (right), corrected for their secular declines ($\dot{m}_{\rm cor}$). The original (uncorrected) data ($\dot{m}$) are shown as dotted curves, with the corresponding exponential fits over-plotted as black, dashed curves.}
\label{fig:mdot_rin_corr}
\end{figure*}
Among the several models in the literature that try to explain X-ray variability, a common assumption is that the underlying process is stationary, i.e., the mean and variance of the time series do not change over time. However, this is not what we see in our $\dot{m}$ profiles, especially for simulations A and R, which are far from being stationary. Therefore, it may be worthwhile to try to correct for any secular behavior seen in these simulations. Unfortunately, there is no foolproof method for doing so, and any procedure presents the risk of introducing artifacts into the data. Nevertheless, for purposes of comparison, we adapt the method of \citet{RM09} to apply an exponential fit to $\dot{m}$. Since simulation D does not show much of a decline and simulation C instead shows an increase followed by a late decline, we present the corrected accretion rate ($\dot{m}_{\rm cor}$) profiles at $r_{\rm H}$ only for simulations A, B, and R in Fig.~\ref{fig:mdot_rin_corr}, while the raw rates from Fig.~\ref{fig:mdot_rin} are reproduced as dotted curves. The exponential fit used in each case is shown as a black, dashed curve. In the following subsections, we shall use these $\dot{m}_{\rm cor}$ profiles to understand if a secular decline alters any of our results. 

\begin{figure*}
\begin{subfigure}[t]{0.49\textwidth}
         \centering
         \includegraphics[width=\textwidth]{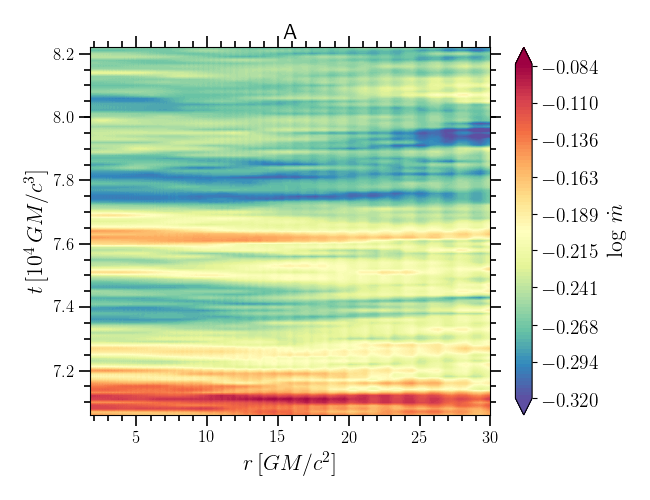}
         \label{fig:mdot_A}
     \end{subfigure}
     \begin{subfigure}[t]{0.49\textwidth}
         \centering
         \includegraphics[width=\textwidth]{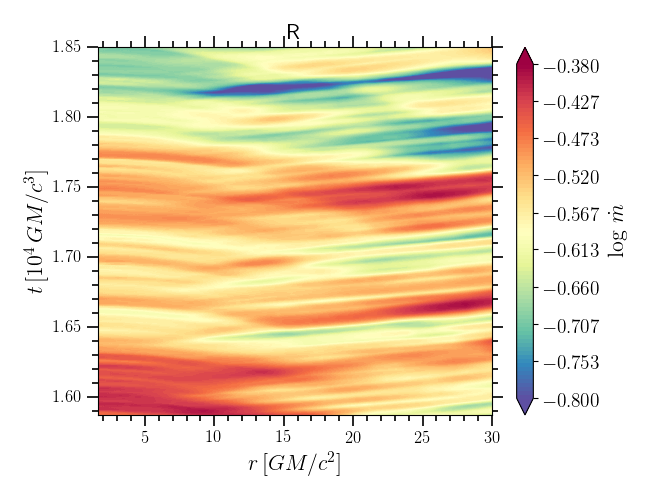}
         \label{fig:mdot_RN}
     \end{subfigure}
\\
     \begin{subfigure}[t]{0.49\textwidth}
         \centering
         \includegraphics[width=\textwidth]{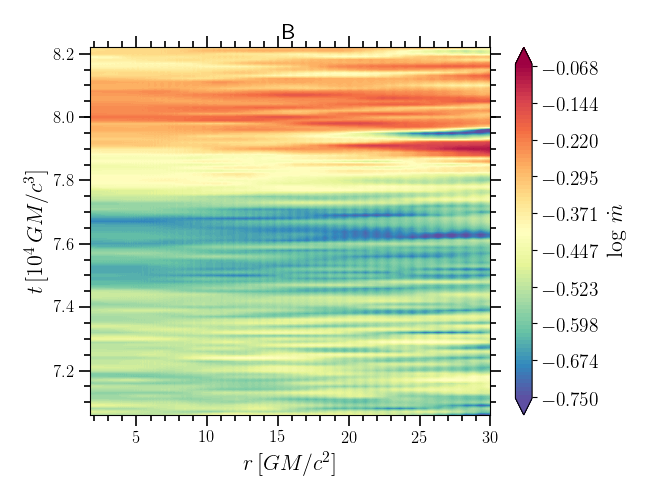}
         \label{fig:mdot_B}
     \end{subfigure}
     \begin{subfigure}[t]{0.49\textwidth}
         \centering
         \includegraphics[width=\textwidth]{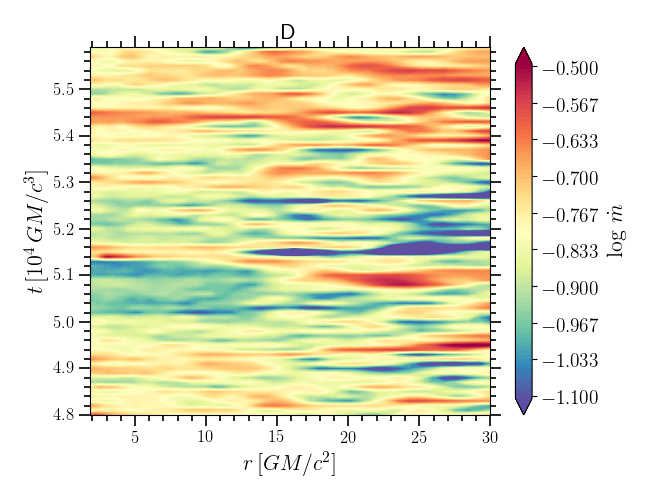}
         \label{fig:mdot_DB}
     \end{subfigure}
\\
\begin{flushleft}
\begin{subfigure}[t]{0.49\textwidth}
 \includegraphics[width=\textwidth]{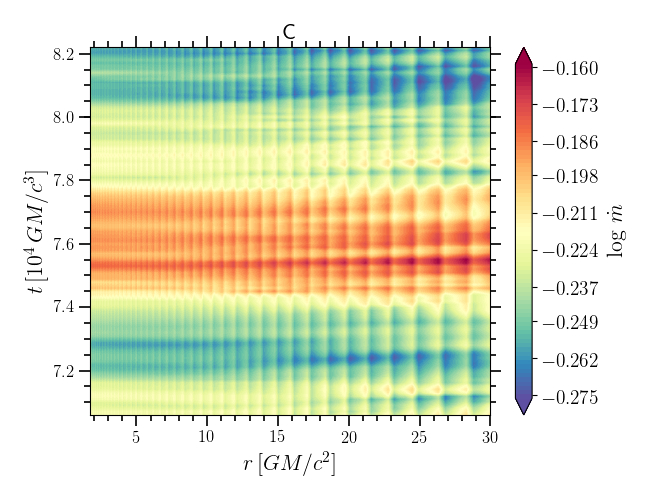}
\label{fig:mdot_C}
\end{subfigure}
\end{flushleft}
\caption{Spacetime diagrams of $\dot{m}$ for all five simulations from $t=30\,P_{\rm orb}$ to $35\,P_{\rm orb}$. Notice the color scale is different for each panel. The faint vertical stripes seen in simulations A, B and C are artifacts of extracting $\dot{m}$ over multiple refinement layers.}
\label{fig:mdot_spacetime}
\end{figure*}

In Fig.~\ref{fig:mdot_spacetime}, we show spacetime plots of $\dot{m}$ for each simulation after $30\,P_{\rm orb}$ over a window of $5\,P_{\rm orb}$. Over this time window, the simulations are all in inflow equilibrium within $30\rg$. Simulations A, B, and C show lower overall $\dot{m}$ during this period and lower variability. Simulation R shows an interesting chevron pattern, which indicates that there are $\dot{m}$ changes originating between $r=10$ and $15\rg$ that are then propagating inwards {\it and} outwards. Simulation D is distinguished by exhibiting the most small-scale variability. 

\section{Results}
\label{sec:results}
\subsection{Power-spectra}
\label{sec:psd}
We primarily use power spectra to look for traces of variability in these simulations. The power spectra are computed using the normalised $\dot{m}$ for each simulation following the steps presented in \citet{UCFKW14}. At each radius, the time series data, i.e., the normalised accretion rate, is binned into a certain number of segments with the requirement that the number of time intervals per segment ($N$) be an integral power of 2. For each segment, a  normalised periodogram, $P$, is computed:
\begin{equation}
    P = \frac{2\Delta t}{\langle \dot{m} \rangle^2 N}|\widetilde{\dot{m}}(\nu)|^2 ~,
\end{equation}
where $\widetilde{\dot{m}}(\nu)$ is the discrete Fourier transform of $\dot{m}$ \citep[defined as in eq.~1 of][]{UCFKW14} and the angle brackets denote a time-average value. We compute $P$ below the maximum frequency of $1/(2 \Delta t)$ and above a minimum frequency given by the inverse of the segment length. Here $\Delta t$ is the sampling time of the data, i.e., the dumping interval from the original simulation (provided in Table~\ref{tab:sim_info}). The average of the periodograms obtained from all the segments yields the average power spectral density (PSD). To further reduce the noise, the resulting power spectrum from the previous procedure is re-binned logarithmically over frequencies and averaged over each frequency bin. This gives the normalised root mean square PSD, which is what we plot in all the power spectra figures. The frequencies in all the power spectra are reported in Hz, assuming a unit solar mass black hole; therefore, by simply dividing by the mass in solar units, the frequencies for any other black hole mass can be extracted. 

For all the space-frequency plots discussed in this section, we also plot three characteristic frequencies of disc dynamics: the Keplerian, radial epicyclic and viscous frequencies. The viscous frequency ($\nu_{\rm visc}$) is simply computed as \begin{equation}
    \nu_{\rm visc}(r) = \dfrac{\langle V^r(r,t) \rangle}{r} ~,
    \label{eq:visc}
\end{equation}
so it is directly related to the accretion time-scale, where $V^{r}$ is the density-weighted, radial component of the 3-velocity averaged over poloidal and azimuthal angles within one scale-height of the mid-plane.

For all the simulations, we also extract the power spectra at 2, 10 and $20\rg$. Of course, we certainly would not expect significant radiation to come from $r=2r_g$, but as Fig.~\ref{fig:mdot_spacetime} shows, the mass flux is nearly uniform for all simulations at radii less than the innermost stable circular orbit (ISCO; at $4.23\rg$ for simulation D and $6\rg$ for the rest). This indifference to the presence of an ISCO is a common feature of ADAFs. Thus, the choice to consider $\dot{m}$ at radii below the ISCO should not adversely affect our variability analysis. For each of the resulting power spectra, we do a least-square fitting with either a single power-law, $P \propto \nu^{\beta}$, or a broken power law, with $P\,\propto\nu^{\beta_1}$ for $\nu < \nu_{\rm break}$ (break frequency) and $P\,\propto \nu^{\beta_2}$ for $\nu> \nu_{\rm break}$. In almost all cases, we find that the power spectra of $2\rg$ and $10\rg$ are highly similar below the viscous frequency of $10\rg$ and so we present the power-law fittings\footnote{Note that, although we plot $P\nu$ in the spectral plot, the respective power law indices labeled in the plots correspond to the fitting made to $P$ and not $P\nu$.} to the averaged PSD of $2\rg$ and $10\rg$. 

\subsubsection{\textbf{Simulation A}}
\begin{figure}
 \centering
\includegraphics[scale=0.52]{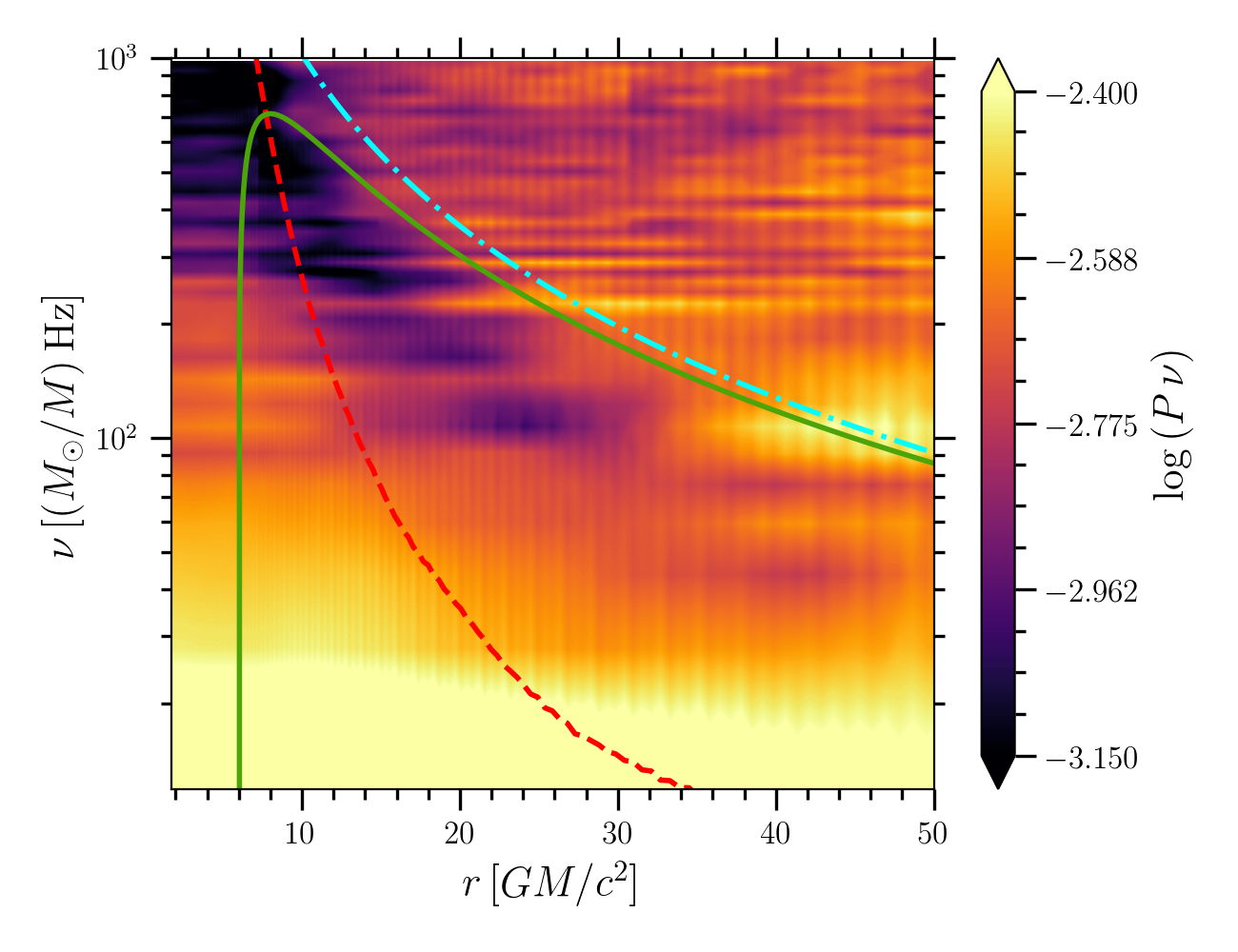}
\caption{Normalised power spectra of $\dot{m}$ for simulation A computed for the time chunk $[38900, 440000]\,GM/c^3$, binned into segments of length $0.13\,(M/M_{\odot})$~s. The blue, dash-dot; solid, green; and red, dashed curves show the Keplerian, radial epicyclic and viscous frequencies, respectively.}
\label{fig:psdspacetime_A}
\end{figure}

Fig.~\ref{fig:psdspacetime_A} shows the power spectra of $\dot{m}$ from simulation~A computed over a time window $[38900,~440000]\,GM/c^3$, during which inflow equilibrium is established out to $100\rg$. Each segmented bin is of the length $0.13\,(M/M_{\odot})$~s. It is interesting to note the power beyond the radial epicyclic frequency curve\footnote{Hereafter, by beyond the radial epicyclic curve, we mean at frequencies above the radial epicyclic curve for radii larger than where that curve peaks.} and within the viscous frequency curve. Between these curves, we find significantly less power. The power above the radial epicyclic frequency is possibly due to pressure waves, analogous to discoseismic $p$-modes, which are likely non-dissipative and therefore will not contribute to the observed flux variations \citep{NK2009}. Power below
the viscous frequency is likely due to propagating fluctuations. More evidence for the second statement will be provided in the later subsections when we discuss radial coherence and time lags. 
Below the epicyclic curve, as the frequency increases, the power drops and reaches a broad minimum at a factor of a few below the Keplerian frequency. This is not a true break in the power spectrum, as the power recovers at larger frequencies. However, it is interesting to note that in the case of geometrically thin discs, a break frequency has been predicted to correspond to the dynamo time-scale~\citep{King2004}, which is a few times the Keplerian time-scale. This was later confirmed by the simulations of \citet{HR16} and may be related to what we are seeing here. A similar break, happening below the local Keplerian frequency, has recently been inferred from X-ray pulsar data \citep{Mushtukov19}.

Since $\dot{m}$ does not vary noticeably across the ISCO, as pointed out in the previous section, it is not surprising that the power spectra here and in subsequent sections also do not bear any signature of it. This in contrast with thin disc simulations, where the nature of the variability changes dramatically at the ISCO \citep{RM09, MWF2019}.

\begin{figure*}
\begin{subfigure}[t]{0.49\textwidth}
         \centering
         \includegraphics[width=\textwidth]{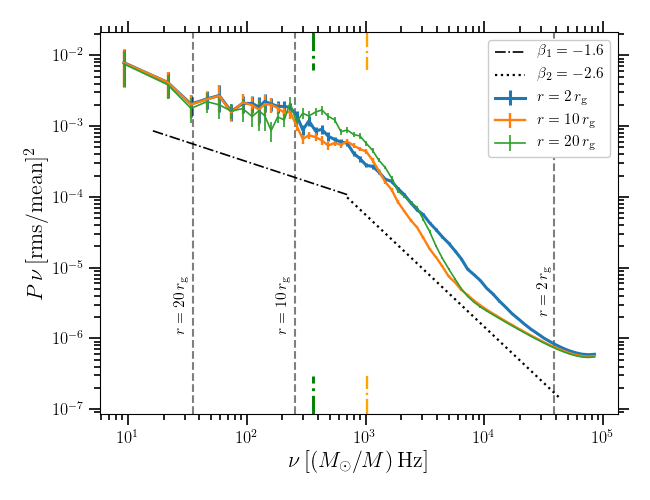}
     \end{subfigure}
     \begin{subfigure}[t]{0.49\textwidth}
         \centering
         \includegraphics[width=\textwidth]{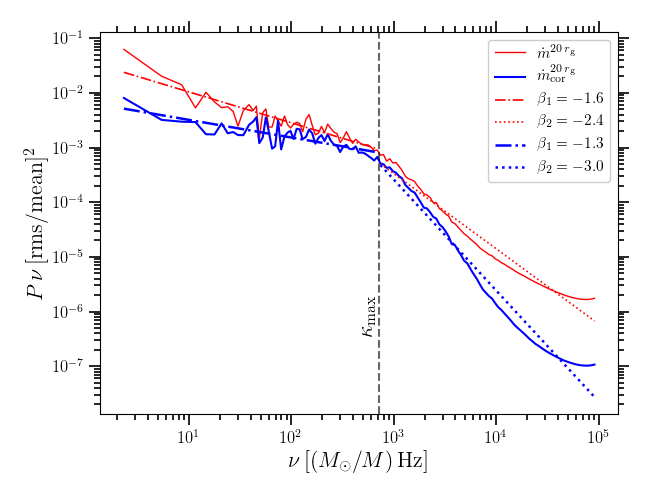}
     \end{subfigure}
\caption{Normalised power spectra of $\dot{m}$ for simulation A at selected radii using the high cadence data ($\Delta t=1\,GM/c^3$) over a time window of $[38900, 440000]\,GM/c^3$. \textit{Left panel}: PSDs for radii $r=2,\,10$ and $20\rg$ computed using bins of size $0.16\,(M/M_{\odot})$~s, shown in blue, orange and green curves, respectively, with decreasing line thickness. The average slopes of the broken power law fittings to all three curves are indicated with black, dash-dot and dotted lines. The Keplerian frequencies for $10$ and $20\rg$ are marked near the $x$-axes in thin, orange and thick, green dash-dot lines, respectively. The viscous frequencies for all three radii are shown as vertical dashed lines. \textit{Right panel}: Average of the normalised PSDs for radii $2,\, 5,\,10$ and $20\rg$ using longer bins of length $0.65\,(M/M_{\odot})$~s shown in thin, red and thick, blue curves for the original $\dot{m}$ and corrected $\dot{m}_{\rm cor}$ data. The respective broken power law fittings are shown in dash-dot and dotted lines. The vertical, dashed line shows the maximum of the radial epicyclic frequency, which occurs at $8\rg$.}
\label{fig:psdradii_A}
\end{figure*}

In the left panel of Fig.~\ref{fig:psdradii_A} we extract the power spectra at radii $r=2,\,10$, and $20\rg$, shown as blue, orange and green solid curves with decreasing line thickness, respectively. The chosen bin length is close to $0.16\,(M/M_{\odot})$~s. For both the plots in Fig.~\ref{fig:psdradii_A}, we use the high cadence data with $\Delta t=1\,GM/c^3$, which is available only for simulation~A and only for radii $r= 2, 5, 10$ and $20\rg$. The PSDs are well fit by a broken power law with an index close to $-1.55$ for low frequencies, and an index $\sim -2.57$ for high frequencies. The break occurs at around $10^3(M_{\odot}/M)\,$Hz, which happens to be where Fig.~\ref{fig:psdspacetime_A} cuts off. The low-frequency slope is slightly steeper than flicker-type noise (index $\approx -1$), and the high-frequency slope is slightly steeper than red-noise behaviour (index $\approx -2$).

If we study Fig.~\ref{fig:psdradii_A} closely, we see that the spectrum at each radius shows strong correlation with the spectra of smaller radii below the local viscous frequency. For example, the curves of power spectra at 2, 10 and $20\rg$ are virtually the same below the dashed line corresponding to the viscous frequency at $20\rg$. Similarly the PSD amplitudes of $2\rg$ and $10\rg$ are nearly identical below the dashed line showing the viscous frequency at $10\rg$. This seems to disagree with the assumption of the propagating fluctuations model that each independent annulus produces fluctuations with most of the variability power centred at the local viscous frequency. Moreover, it rules out the damping of propagating fluctuations \citep{AU06, RIV2017, Mahmoud18b}.
Fig.~\ref{fig:psdradii_A} also shows that the break frequencies of the spectra do not correspond directly to any of the characteristic frequencies we consider, but do lie close to the local Keplerian frequencies.

Since the observed flux is usually an integration of the emission from a range of disc radii, we compute the average of the PSD at radii $2, 5, 10$ and $20\rg$, shown as the thin, red curve in the right panel of Fig.~\ref{fig:psdradii_A}. The broken power law fits are shown in thin, red dash-dot and dotted lines, respectively. We then checked if the results differ when we use the accretion rate corrected for the secular decline. The thick blue curve in the same plot shows the average PSD computed for the same radii but now using $\dot{m}_\mathrm{cor}$ data. The segment length used in this plot is $0.65\,(M/M_{\odot})$~s, and so the power spectrum extends down to $1.5\,(M_{\odot}/M)$~Hz, where we see a significant difference between the PSD of $\dot{m}$ and $\dot{m}_\mathrm{cor}$. When compared to the left panel plot, which has shorter segment length (i.e., $0.16\,(M/M_{\odot})$~s), the PSD of $\dot{m}_\mathrm{cor}$ coincides with the PSD of the original (uncorrected) $\dot{m}$ at all the frequencies except the high frequency tail, which starts to differ above $\approx 3000\,(M_{\odot}/M)$~Hz. Thus the difference in the PSDs from the original $\dot{m}$ to the corrected $\dot{m}_{\rm cor}$ data starts to become significant as the length of the binned segments increases. Perhaps, this is because the longer time-period variations imposed by the secular decline in $\dot{m}$ can be traced better if the segments are longer, which leads to dominant power at low frequencies in the power spectra.

The break frequencies of the average PSD in the right panel of Fig.~\ref{fig:psdradii_A} do not correspond to the viscous nor the Keplerian frequencies of any of the four radii (2, 5, 10 and $20\rg$). Neither do they correspond to any of the characteristic frequencies of the inflow-equilibrium radius. Instead, they appear to be more related to the peak of the radial epicyclic frequency curve, $\kappa_\mathrm{max}$.

\subsubsection{\textbf{Simulation B}}
\begin{figure*}
\begin{subfigure}[t]{0.49\textwidth}
         \centering
         \includegraphics[width=\textwidth]{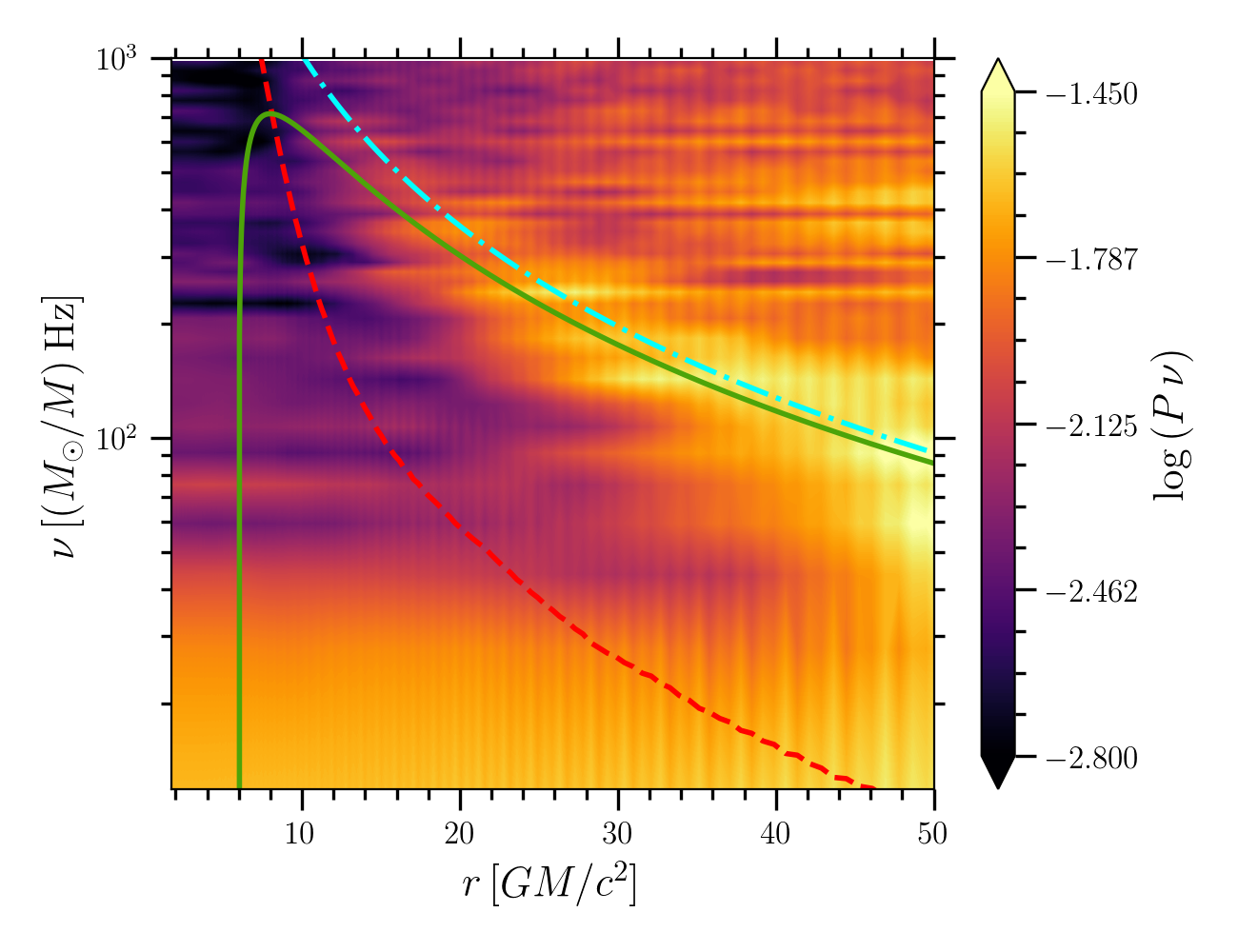}
     \end{subfigure}
         \centering
     \begin{subfigure}[t]{0.49\textwidth}
         \includegraphics[width=\textwidth]{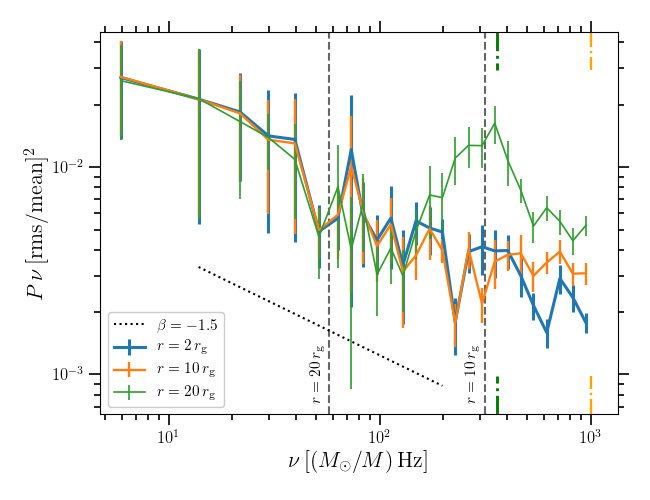}
     \end{subfigure}
\caption{Normalised power spectra of $\dot{m}$ for simulation~B. \textit{Left panel}: Variation of power with frequency and radius computed over the time period $[38900, 220000]\,GM/c^3$ using bins of length $0.13\,(M/M_{\odot})$~s. Blue, dash-dot; solid, green; and red, dashed curves show the Keplerian, radial epicyclic and viscous frequencies, respectively. \textit{Right panel}: PSDs at 2, 10 and $20\rg$ for the same time period but with longer segment lengths of $0.25\,(M/M_{\odot})$~s. The average slope of the single power law fittings to the 2 and $10\rg$ curves is indicated by the dotted line. The black, dashed lines labeled with radii represent the local viscous frequencies, while the thin, orange and thick, green, dash-dot lines near the $x$-axes mark the Keplerian frequencies at 10 and $20\rg$, respectively.}
\label{fig:psdB}
\end{figure*}

The left panel in Fig.~\ref{fig:psdB} shows the power spectra for simulation B, for radii within $50\rg$, computed over the time period $[38900, 220000]\,GM/c^3$ binned into segments of length $0.13\,(M/M_{\odot})$~s (same as for simulation~A). There is definitely more power above the radial epicyclic frequency, but power within the viscous frequency curve is not as prevalent as in simulation~A. In this particular simulation, power along the radial epicyclic and Keplerian frequency curves is more evident, and it seems to be present throughout the simulation period. 

The usual interpretation of observations put forward by the propagating fluctuation model requires that at higher frequencies, smaller radii should exhibit more power than larger radii. The right panel of Fig.~\ref{fig:psdB} seems to contradict this. In the figure, we plot the PSDs for the same radii ($r=2,\,10$ and $20\rg$) as Fig.~\ref{fig:psdradii_A} with a longer segment length of $0.25\,(M/M_{\odot})$~s. Clearly, the PSD at $20\rg$ has more power at higher frequencies compared to $2\rg$ and $10\rg$. This excess is coming from the power present around the radial epicyclic and Keplerian frequencies seen in the left panel (note that the peak of the green curve in the right panel coincides with the Keplerian frequency at $20\rg$ marked by the thick, green, dash-dot line). 

We do not see any clear break in any of the power spectra in the right panel of Fig.~\ref{fig:psdB}, most likely because we are not going to high enough frequencies due to the lower sampling rate of this data. A single power law fit, with an index close to $-1.5$ (black, dashed line), matches the 2 and $10\rg$ curves fairly well. Since $\dot{m}$ does not decrease significantly for simulation~B, the PSD of $\dot{m}_\mathrm{cor}$ is similar to the PSD of $\dot{m}$, and hence we do not show it here.

\subsubsection{\textbf{Simulation C}}
\begin{figure*}
\begin{subfigure}[t]{0.49\textwidth}
         \centering
         \includegraphics[width=\textwidth]{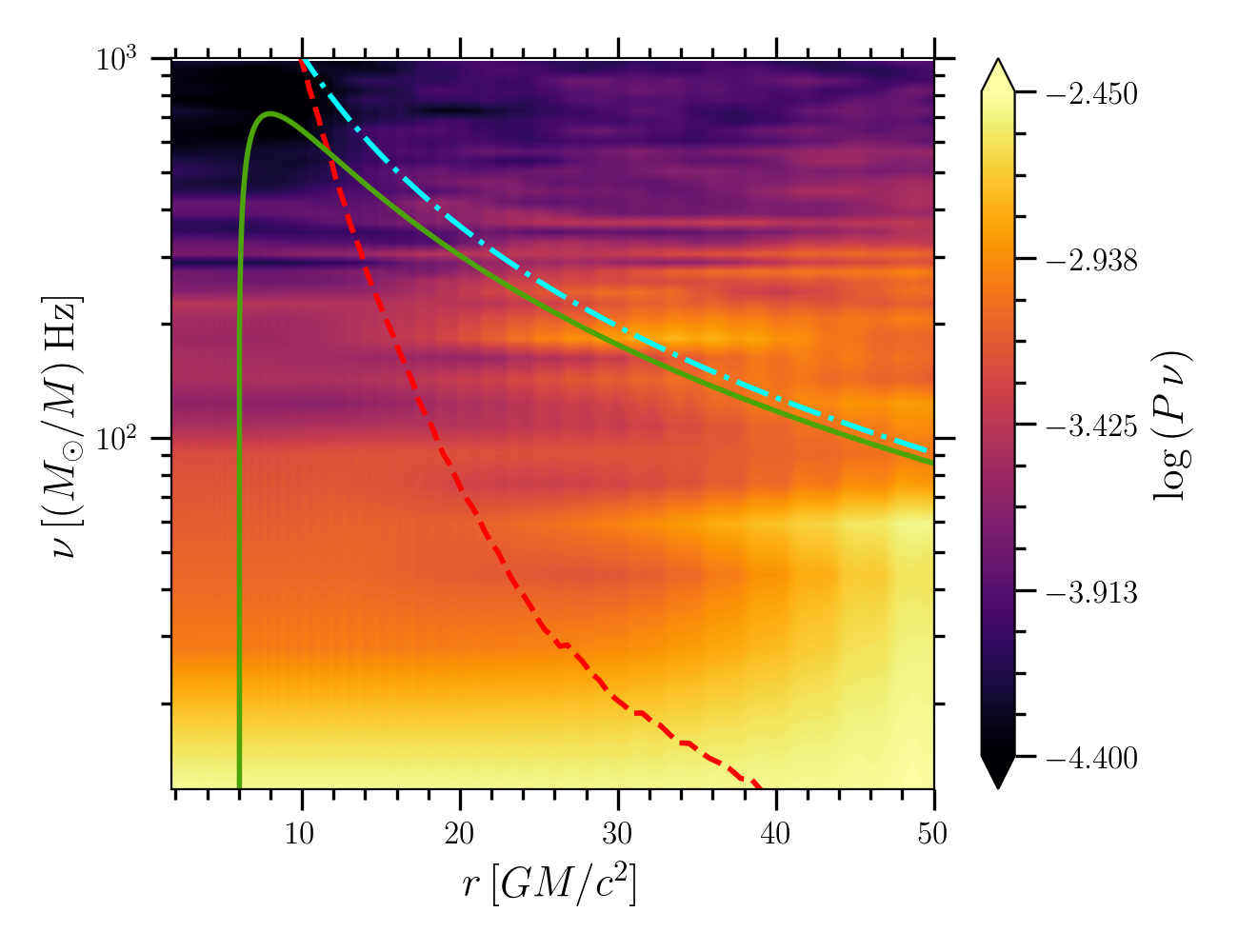}
     \end{subfigure}
         \centering
     \begin{subfigure}[t]{0.49\textwidth}
         \includegraphics[width=\textwidth]{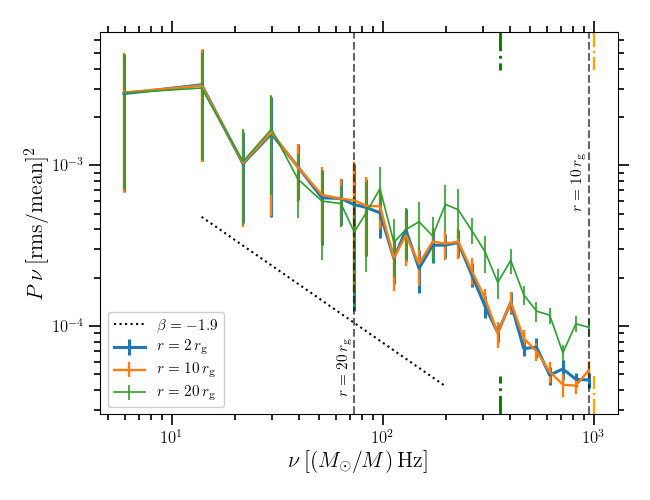}
     \end{subfigure}
\caption{Normalised power spectra of $\dot{m}$ for simulation~C. \textit{Left panel}: Variation of power with frequency and radius computed over the time period $[38900, 220000]\,GM/c^3$ using bins of length $0.13\,(M/M_{\odot})$~s. Blue, dash-dot; solid, green; and red, dashed curves show the Keplerian, radial epicyclic and viscous frequencies, respectively. \textit{Right panel}: PSDs at 2, 10 and $20\rg$ for the same time period but with longer segment lengths of $0.25\,(M/M_{\odot})$~s. The average slope of the single power law fittings to all three curves is indicated by the dotted line. The black, dashed lines labeled with radii represent the local viscous frequencies and the thin, orange and thick, green, dash-dot lines near the $x$-axes mark the Keplerian frequencies at 10 and $20\rg$, respectively.}
\label{fig:psdC}
\end{figure*}

We do a similar analysis for simulation~C and the results are shown in Fig.~\ref{fig:psdC}. The left panel shows the power spectra for radii within $50\rg$ using the same segment length, i.e., $0.13\,(M/M_{\odot})$~s, and over the same time window $[38900, 220000]\,GM/c^3$ as simulation B. Unlike the previous two simulations, simulation~C does not seem to exhibit a strong preference for power above the radial epicyclic frequency. There is also no dominant power along the radial epicyclic frequency/Keplerian frequency curve, nor is there much distinction between inside and outside the viscous frequency curve. This agrees with Fig.~\ref{fig:mdot_spacetime}, where we see that simulation~C shows less rapid variability compared to the other simulations. 

In the right panel, we plot the PSDs for the same radii ($r=2,\,10$ and $20\rg$) with the longer segment length of $0.25\,(M/M_{\odot})$~s. Similar to simulation B, we see more high-frequency power at large radii than small and see no evidence for a break in the power spectrum, likely for the same reason as for simulation~B. A single power law fit to the power spectra is provided (black, dotted line), with an index close to $-1.91$, which is perhaps consistent with red-noise behaviour. 

\subsubsection{\textbf{Simulation R}}
\begin{figure}
\centering
\includegraphics[scale=0.52]{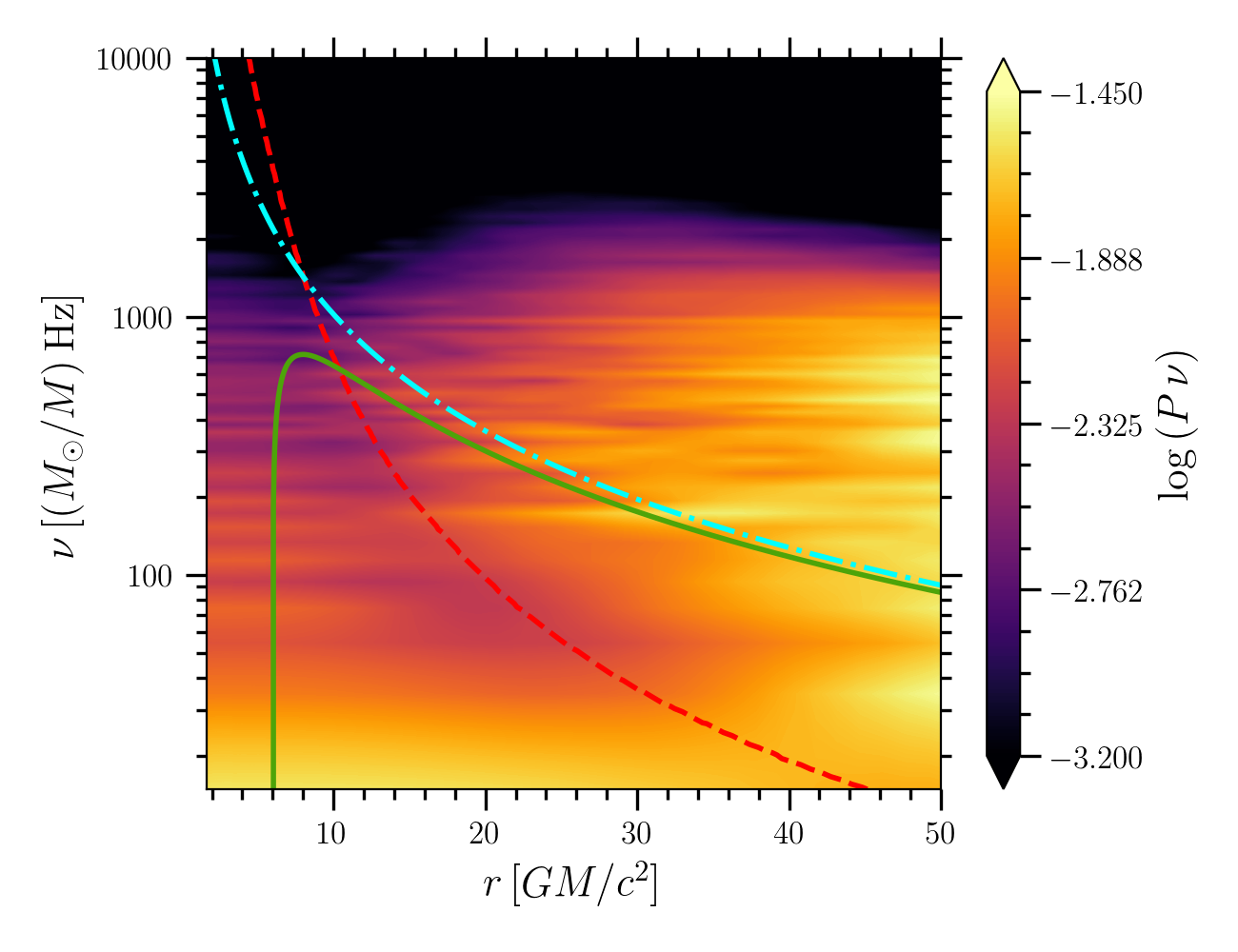}
\caption{Normalised power spectra of $\dot{m}$ for simulation R computed for the time period $[50000, 236670]\,GM/c^3$ binned into segments of length $0.11\,(M/M_{\odot})$~s. The blue, dash-dot; solid, green; and red, dashed curves show the Keplerian, radial epicyclic and viscous frequencies, respectively. }
\label{fig:psdR_spacetime}
\end{figure}

Simulation~R has a higher sampling frequency ($\Delta t=10\,GM/c^3$), which gives the advantage of being able to study higher frequencies and also obtain better spectra as each bin will now have more data points to average out the noise. In Fig.~\ref{fig:psdR_spacetime}, we show the power spectra of $\dot{m}$ computed over the time period $[50000, 236670]\,GM/c^3$, during which the flow has reached inflow equilibrium to beyond $50\rg$ \citep[see table 1 and fig. 6 of][]{NSPK2012}, with a segment length of $0.11\,(M/M_{\odot})$~s. Interestingly, simulation~R exhibits dominant power above the viscous frequency and does not show any difference in power with respect to the radial epicyclic nor Keplerian frequencies. It is worth noting that the power drops drastically at higher frequencies, roughly at the maximum radial epicyclic frequency. 

\begin{figure*}
         \centering
     \begin{subfigure}[t]{0.49\textwidth}
         \includegraphics[width=\textwidth]{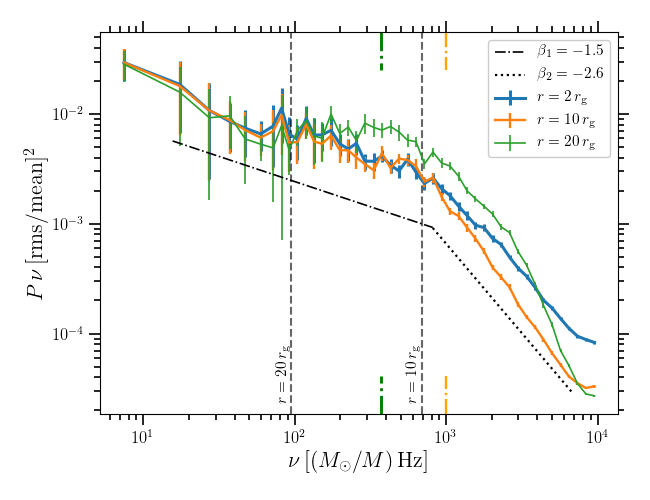}
     \end{subfigure}
        \centering
        \begin{subfigure}[t]{0.49\textwidth}
         \includegraphics[width=\textwidth]{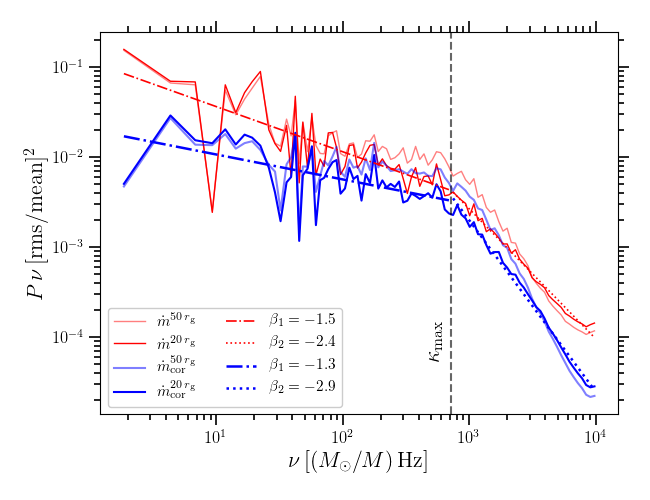}
     \end{subfigure}
\caption{Normalised power spectra of $\dot{m}$ for simulation R at chosen radii. \textit{Left panel}: PSDs computed from the time window $[12000,236000]\,GM/c^3$ using bins of length $0.2\,(M/M_{\odot})$~s are shown in blue, orange and green curves with decreasing line thickness for radii $r=2,\,10$ and $20\rg$, respectively. The average slopes of the power law fittings to the 2 and $10\rg$ curves are indicated by the black dash-dot and dotted lines. The Keplerian frequencies at $10$ and $20\rg$ are marked near the $x$-axes in thin, orange and thick, green, dash-dot lines. The viscous frequencies for all three radii are shown with vertical dashed lines. \textit{Right panel}: Average of the PSDs from all radii within $20\rg$ for the original $\dot{m}$ and corrected $\dot{m}_{\rm cor}$ data using longer bins of $0.8\,(M/M_{\odot})$~s are shown in thin, red and thick, blue curves, respectively. Similar curves for averaged PSD within $50\rg$ are shown in lighter shades. The broken power-law fittings for the $20\rg$ averaged data are shown in the dash-dot and dotted lines. The vertical, dashed line shows the maximum of the radial epicyclic frequency, which occurs at $8\rg$.}
\label{fig:psdR_radii}
\end{figure*}

In Fig.~\ref{fig:psdR_radii}, we show the power spectra extracted at different radii. The left panel shows the spectra at radii 2, 10 and $20\rg$ computed using segments of $0.2\,(M/M_{\odot})$~s over the time window\footnote{By $12000\,GM/c^3$, inflow equilibrium is established up to a radius of $20\rg$.} $[12000, 236670]\,GM/c^3$.
As in simulation~A, we can clearly see that the power spectrum at a given radius is highly coherent with the spectra at smaller radii below the local viscous frequency. This is the case in the other simulations (B, C and D) as well, but it is harder to notice with the power spectra limited to a small Fourier frequency range due to the relatively poor sampling frequency. Simulation R spectra exhibit a clear break in the frequency, at about $800\,(M_{\odot}/M)$~Hz, where the power law index changes from $\sim\,-1.46$ at low frequencies to $\sim\,-2.63$ at high frequencies.

Similar to simulation A, we show a comparison of the radially averaged power spectra for $\dot{m}$ and $\dot{m}_\mathrm{cor}$ in the right panel of  Fig.~\ref{fig:psdR_radii} using segments of length $0.8\,(M/M_{\odot})$~s. The thin, red curves are the spectra  obtained from averaging the PSD of $\dot{m}$ at all radii within 20 and $50\rg$. The thick, blue curves represent the same for $\dot{m}_{\rm cor}$. Power-law fittings for the averaged spectra within $20\rg$ are included. The spectra from $\dot{m}_{\rm cor}$ have a slope that is 15 percent flatter at low frequencies compared to the spectra from the original $\dot{m}$. Similar to the average spectra from simulation A, the frequency break in the right panel of Fig.~\ref{fig:psdR_radii} occurs very close to the maximum radial epicyclic frequency. As we noted earlier, this is attributable to the steep decrease in power above roughly $1\,(M_{\odot}/M)$~kHz.

\subsubsection{\textbf{Simulation D}}
\begin{figure*}
\begin{subfigure}[t]{0.49\textwidth}
         \centering
         \includegraphics[width=\textwidth]{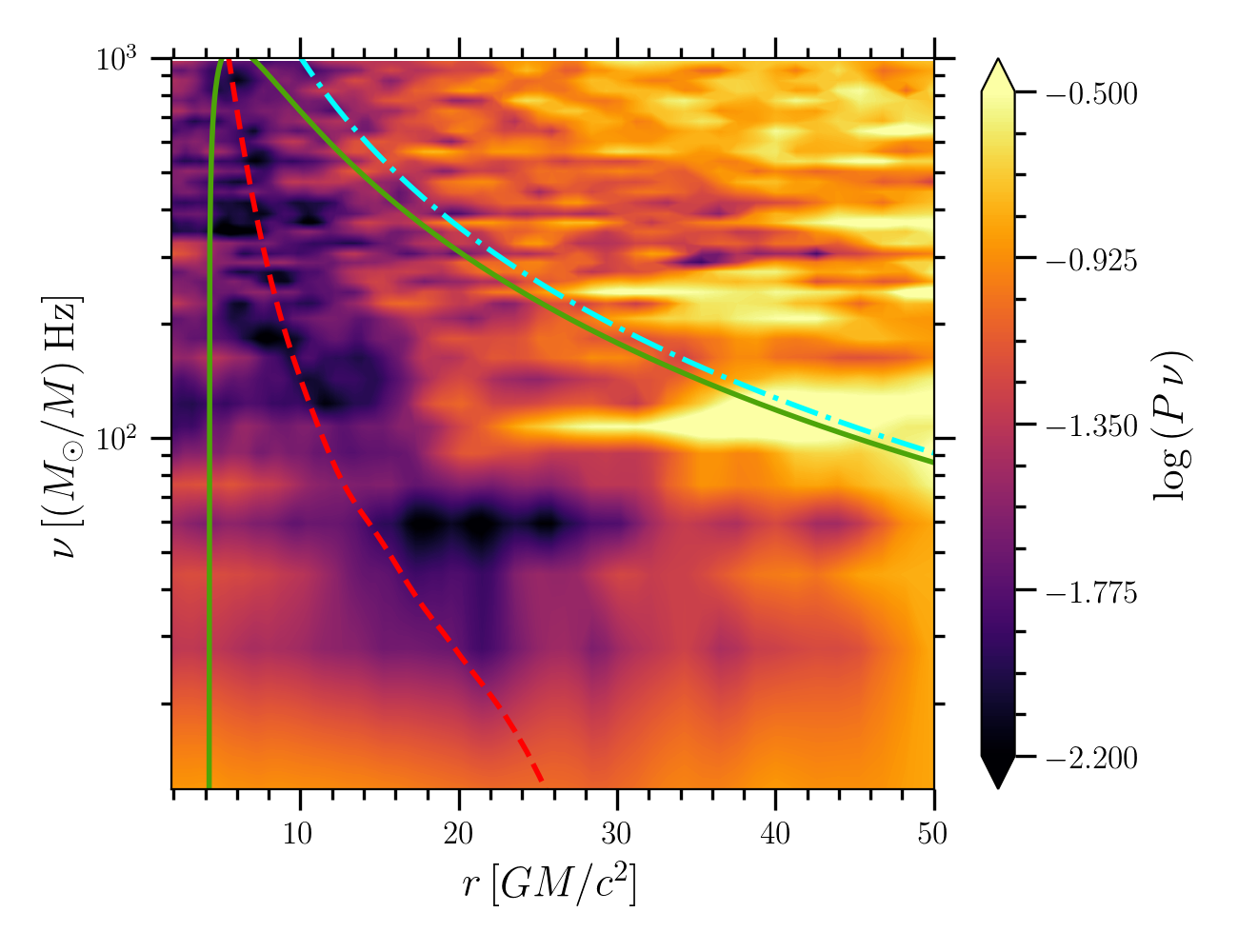}
     \end{subfigure}
         \centering
     \begin{subfigure}[t]{0.49\textwidth}
         \includegraphics[width=\textwidth]{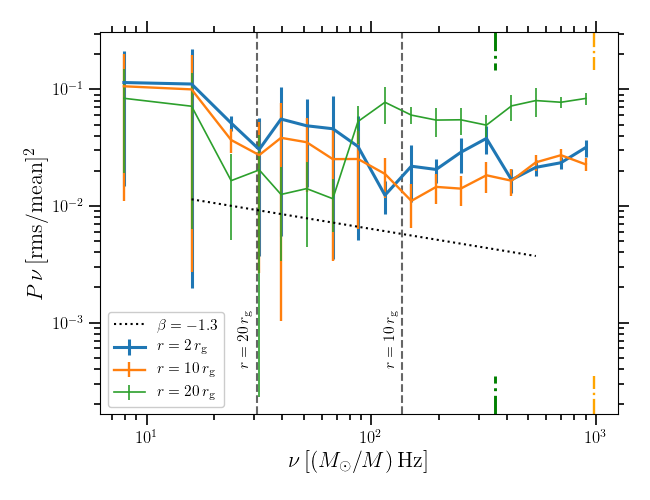}
      \end{subfigure}
\caption{Power spectra of $\dot{m}$ for simulation D. \textit{Left panel}: Normalised PSD computed for the data in the time window $[11000, 63000]\,GM/c^3$ using segments of length $0.13\,(M/M_{\odot})$~s. \textit{Right panel}: Plot of radial slices at $r=2,\,10$, and $20\rg$. The dotted line shows the average slope obtained from averaging the power law fits for $2\rg$ and $10\rg$ curves. Vertical dashed lines give the viscous frequency for labelled radii, and the thin, orange and thick, green, dash-dot lines close to the $x$-axis mark the Keplerian frequencies for $10\rg$ and $20\rg$, respectively.}
\label{fig:psdD}
\end{figure*}

Simulation~D is run for a shorter duration, so the power spectra are computed from the averaged periodograms of only two bins with segment lengths close to $0.13\,(M/M_{\odot})$~s. We use the $\dot{m}$ data over the time window $[11000, 63000]\,GM/c^3$. The left panel of Fig.~\ref{fig:psdD} shows the frequency-radius variation of power spectra and the right panel shows the spectra for the same binning at radii 2, 10 and $20\rg$. Even with less data, we can see that there is dominant power above the radial epicyclic frequency (for $r>8\rg$) and some power within the viscous frequency curve, similar to simulations A and R. The right panel does not show any clear evidence of a break, most likely because of the low sampling frequency of the data. We find the averaged slope of the single power law fits to the $2\rg$ and $10\rg$ curves is around $-1.25$, closer to flicker-noise behaviour, whereas the best fit to the $20\rg$ curve has a slope of $-0.77$. Here, and in Fig.~\ref{fig:psdB} and \ref{fig:psdC}, the large error bars are most likely due to the fact that the number of time intervals ($N$) in each segment is only a couple of hundred, i.e., at least an order of magnitude less than what is possible for simulations A and R (Fig.~\ref{fig:psdradii_A} and \ref{fig:psdR_radii}).

\begin{figure*}
\begin{subfigure}[t]{0.45\textwidth}
         \centering
          \caption*{$\dot{m}(\theta = \pi/2)$}
         \includegraphics[width=\textwidth]{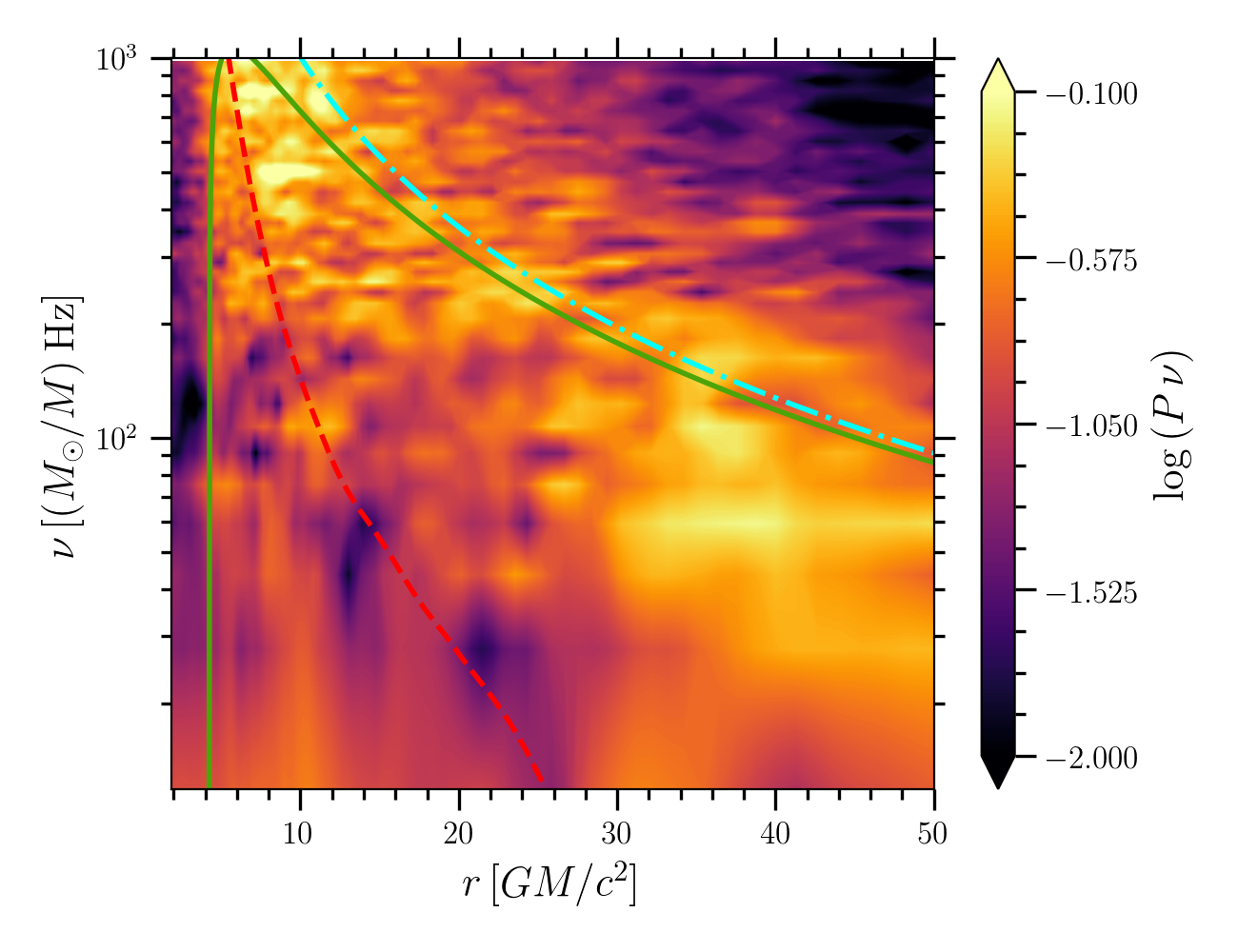}
\end{subfigure}
\begin{subfigure}[t]{0.45\textwidth}
         \centering
          \caption*{$\dot{m}$ within one scale height}
         \includegraphics[width=\textwidth]{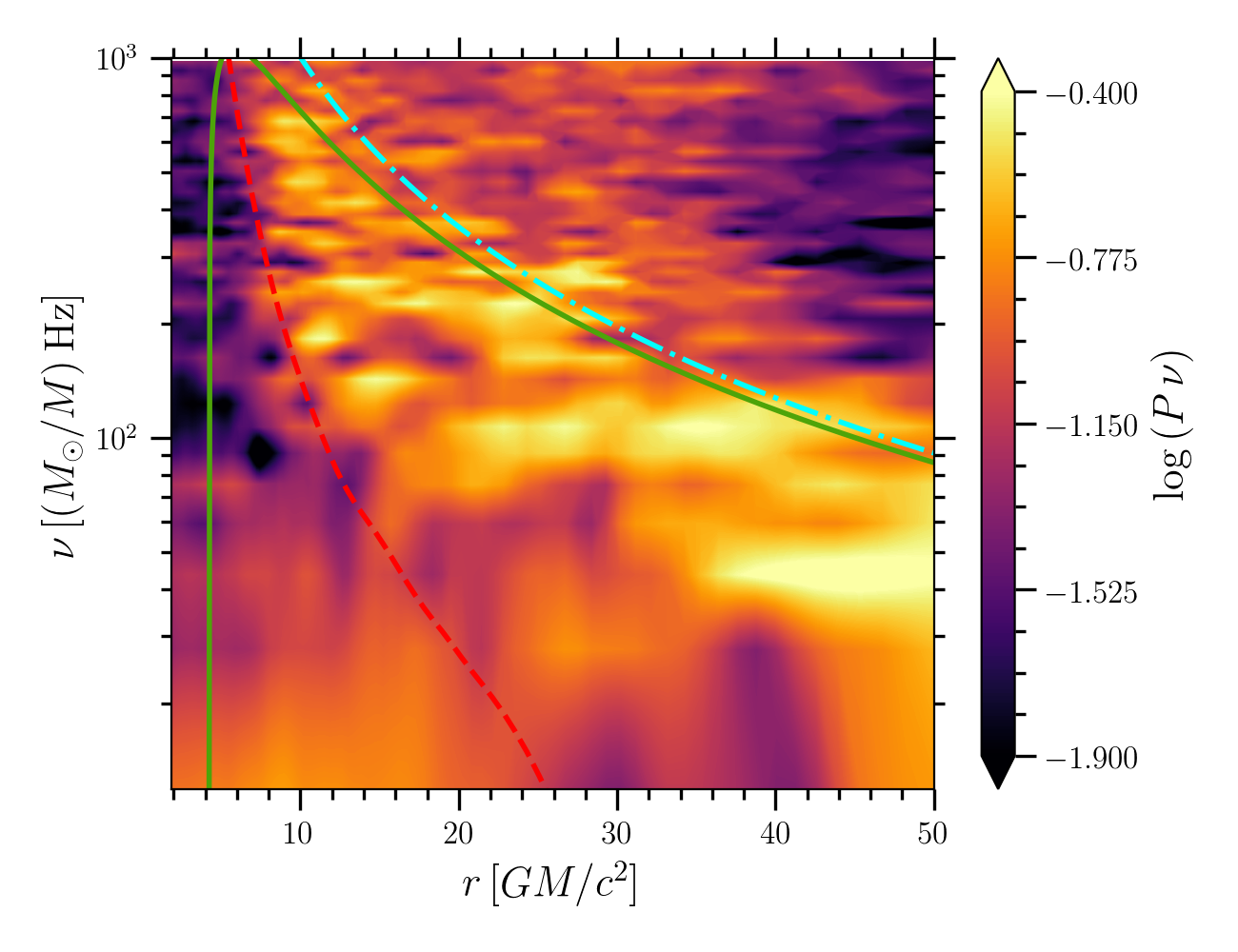}
\end{subfigure}
\caption{Normalised power spectra of the mid-plane value of $\dot{m}$ (left panel) and $\dot{m}$ within one scale height of the midplane (right panel) for simulation D. Spectra are computed over the time window $[11000,63000]\,GM/c^3$ using bins of length $0.13\,(M/M_{\odot})$~s. The blue, dash-dot; solid, green; and red, dashed curves show the Keplerian, radial epicyclic and viscous frequencies, respectively.}
\label{fig:psdmdot}
\end{figure*}

The large amount of power found {\em above} the viscous frequency, and even above the radial epicyclic frequency, in these spectra is certainly not what is expected from the propagating fluctuation model. To explore this further, in Fig. \ref{fig:psdmdot}, we include power spectra from simulation D where we limit the vertical extent of the domain included in the calculation. We consider both the midplane value of $\dot{m}$ (left panel) and $\dot{m}$ within one scale height of the midplane (right panel). In both cases, the relative power found above the radial epicyclic frequency is greatly reduced (compared to Fig. \ref{fig:psdD}). However, most of the power is still above the viscous frequency. 

\subsection{Radial Coherence}
We turn our focus now, though, to the variability power found {\em below} the viscous frequency, the source of which could be the propagating fluctuations we are searching for. According to the propagating fluctuations model, fluctuations driven at different radii should modulate the fluctuations at smaller radii as they propagate in on the viscous time-scale. Any fluctuations above the local viscous frequency should be damped, so they are unable to make it to smaller radii. Thus, the model predicts that the accretion rate between any two radii should exhibit high coherence below the viscous frequency and low coherence above it. In X-ray observations, this translates into a strong coherence between light curves in different energy bands, with the assumption that higher energy bands originate from radii closer to the black hole and lower energy bands come from further out. 

Coherence fundamentally describes the fractional variance between two curves, which can be predicted via a linear transformation between them \citep{VN97}. In our case, let $h(t)$ and $s(t)$ denote $\dot{m}$ curves at any two radii, $r_1$ and $r_2$, with respective Fourier transforms $H(\nu)$ and $S(\nu)$. Then, following \citet{UCFKW14}, we compute the radial coherence of $\dot{m}$ between $r_1$ and $r_2$ using
\begin{equation}
\gamma^2(\nu_j)= \dfrac{|C_{HS}(\nu_j)|^2}{P_H(\nu_j)P_S(\nu_j)} ~,
\label{eq:coherence}
\end{equation}
where $C_{HS} = H^*S$ is the cross-spectrum averaged over multiple time segments and frequency bins and $P_H(\nu_j)$ and $P_S(\nu_j)$ are the power spectra obtained from $h(t)$ and $s(t)$, respectively, with similar averaging. If $h(t)$ and $s(t)$ are related through a linear transformation in time, the two are said to be perfectly coherent, and $\gamma^2$ reaches its maximum value of 1. When $h(t)$ and $s(t)$ are completely unrelated, $\gamma^2=0$ and they are said to be incoherent. 

\begin{figure*}
\centering
\includegraphics[width=\textwidth]{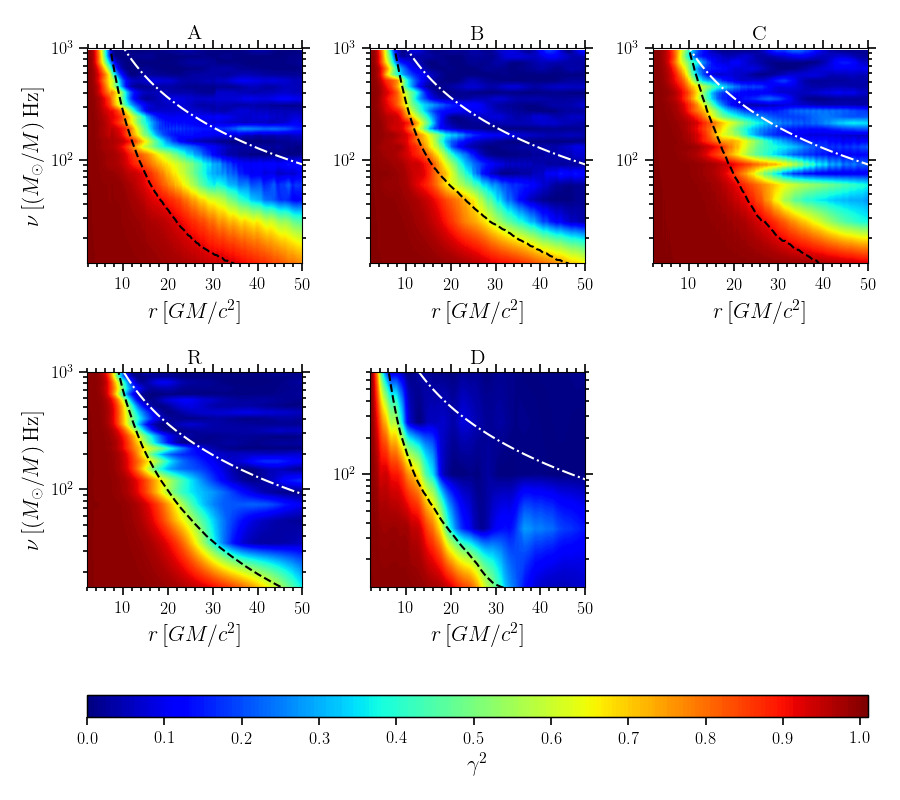}
\caption{Radial coherence measured with respect to the inner radius. The white, dash-dot curve represents the Keplerian frequency and the black, dashed curve is the viscous frequency below which strong coherence is maintained. }
\label{fig:radial_coherence}
\end{figure*}

In Fig.~\ref{fig:radial_coherence}, we show the radial coherence of all radii up to $50\rg$ with respect to the inner radius (black hole horizon) for all five simulations. The black, dashed line represents the viscous frequency, while the white, dash-dot line gives the Keplerian frequency. We consider the same binning of $\dot{m}$ as we did for the radius-frequency plots of power spectra in the previous section. We find that in all cases, $\dot{m}$ shows remarkable coherence below the local viscous frequency at all radii. Similar radial coherence is observed in the simulations and models of geometrically {\it thin} discs~\citep{HR16, Mushtukov18}. In addition, simulations A and C exhibit significant coherence even above the viscous frequency, up to a factor few below the Keplerian frequency, particularly at larger radii. It is interesting to note that these are the two simulations that show equatorial backflow.

\subsection{Time lags}
\label{sec:timelags}

\begin{table}
\begin{center}
\begin{tabular}{c c c c}
\hline
\multicolumn{1}{c}{Sim} & \multicolumn{1}{c}{$r$} & \multicolumn{1}{c}{$t_{\rm acc}$} & \multicolumn{1}{c}{$t_{\rm lag}$} \\
\multicolumn{1}{c}{} & \multicolumn{1}{c}{[$r_g$]} & \multicolumn{1}{c}{[$(M/M_{\odot})$~ms]} & \multicolumn{1}{c}{[$(M/M_{\odot})$~ms]} \vspace{1mm}\\
\hline
\multirow{3}{*}{A} & 2 & 4.57 & 0.25 \\
 & 6 & 4.41 & 0.18 \\
 & 10 & 3.46 & 0.08 \\
 \hline
\multirow{3}{*}{B} & 2 & 3.44 & 0.46 \\
 & 6 & 3.28  & 0.4 \\
 & 10 & 2.52  & 0.25 \\
 \hline
\multirow{3}{*}{C} & 2 & 1.36 & 0.25 \\
 & 6 & 1.26  & 0.18 \\
 & 10 & 0.98 & 0.09 \\
 \hline
\multirow{3}{*}{R} & 2 & 1.65  & 0.44 \\
 & 6 & 1.57 & 0.37 \\
 & 10 & 1.23 & 0.21 \\
 \hline
\multirow{3}{*}{D} & 2 & 7.79 & 1.59 \\
 & 6 & 7.37 & 1.5 \\
 & 10 & 5.32 & 0.88\\
 \hline
\end{tabular}
\end{center}
\caption{Accretion time-scales and time lags for different radii within each simulation. The time lags are averaged from values below the viscous frequency at $15\rg$.}
\label{tab:tlags}
\end{table}

The presence of frequency-dependent, radial coherence indicates that fluctuations at different radii are causally connected. However, information about the direction of propagation of these fluctuations is missing from a simple coherence plot. To extract this information, we compute the time leads/lags between $\dot{m}$ at different radii. The strong coherence seen in Fig.~\ref{fig:radial_coherence} implies that the two mass fluxes must be related by a phase shift ($\Delta \Phi$) that can be calculated from the ratio of the imaginary and real parts of the cross-spectrum, $C_{HS}$. Time lags are then obtained using $\Delta \tau = \Delta \Phi/(2 \pi \nu)$.

\begin{figure*}
\centering
\includegraphics[width=\textwidth]{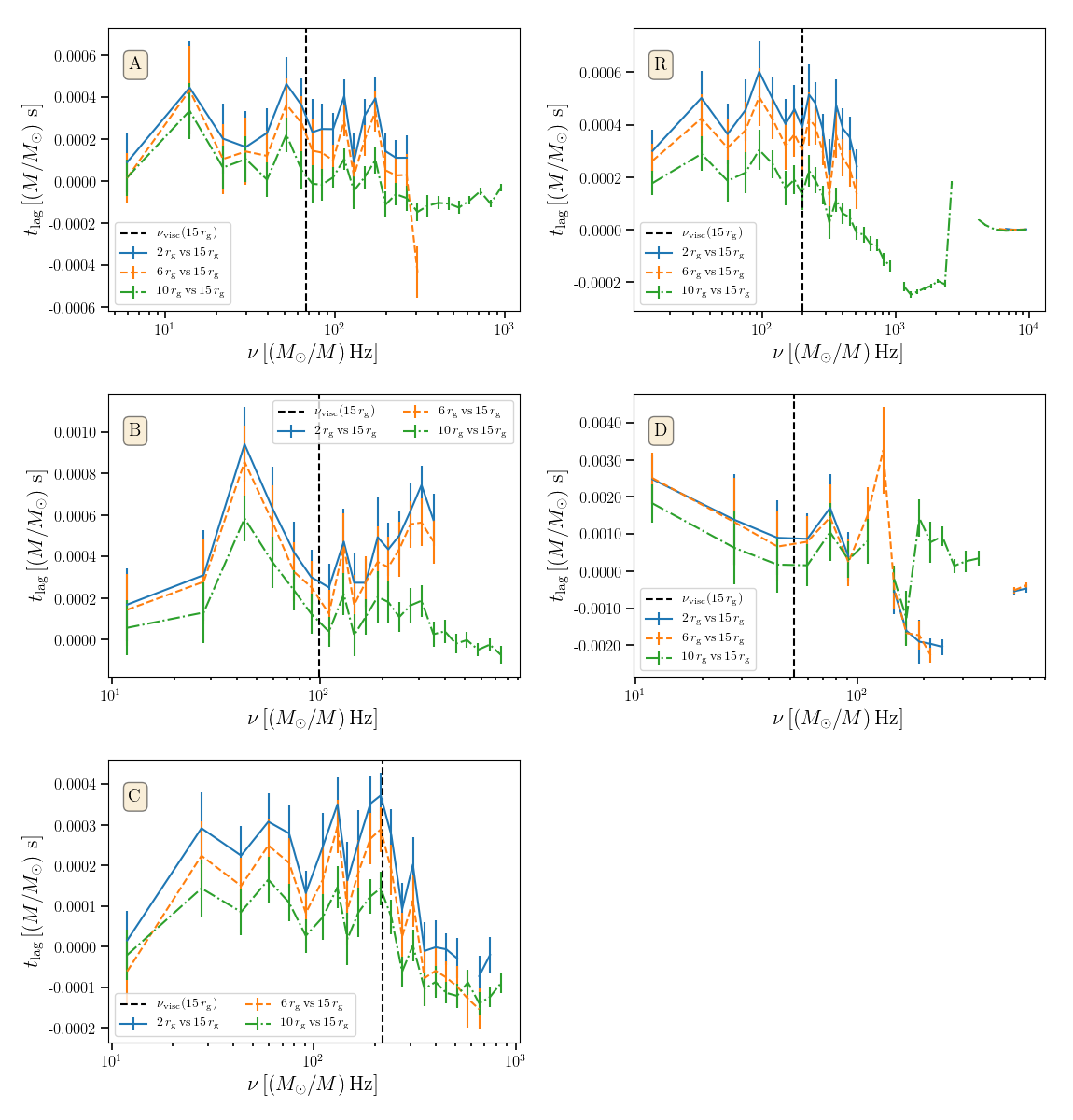}
\caption{Time lags of the $\dot{m}$ fluctuations for radii $2$, $6$ and $10\rg$ with respect to $15\rg$ for all five simulations. The dashed line represents the viscous frequency at $15\rg$. Lags with coherence less than 0.1 are omitted.}
\label{fig:time_lags}
\end{figure*}

In Fig.~\ref{fig:time_lags}, we show the time lags of $\dot{m}$ at $r=2$ (blue, solid curve), 6 (orange, dashed curve), and $10\rg$ (green, dash-dot curve) with respect to $15\rg$. We do not show the lags where the coherence drops below 0.1. We find in all simulations that fluctuations below the viscous frequency at $15\rg$ (black, dashed line) show positive, definite lags when comparing small radii to larger radii, and for the most part, the magnitudes of the lags increase for larger radial separations. However, comparing the measured lags to the local accretion time-scale ($t_\mathrm{acc} = \int_r^{15} V^r(r,t)\,dr$) for each radius (Table~\ref{tab:tlags}), we find they are uniformly shorter, too short to correspond to the viscous time-scale and more consistent with the local Keplerian time-scale. 

We also find some evidence for negative lags at frequencies higher than the viscous frequency. Negative lags are also present in the thin-disc model of \citet[][see figure 9b of their paper]{Mushtukov18}. There, they are explained as high-frequency variability from the inner disc affecting variability at larger radii, due to outward propagating waves; by contrast, inward propagating high-frequency fluctuations from large radii are damped before they reach the inner disc, according to the model. The picture from our thick-disc simulations is less clear, however, as we only see negative lags 
in some of the simulations. In principle, in regions of backflow (which are present in simulations A and C), mass accretion rate fluctuations may also be simply transmitted outwards by the fluid moving away from the black hole. 

One issue with Fig.~\ref{fig:time_lags} is that for frequencies below the viscous one we do not find a well-defined frequency dependence in the time lags. In our analysis, the time lags simply correspond to the propagation times between $15\rg$ and the inner radii, which are independent of the Fourier frequency. It could be that the dissipative processes that convert $\dot{m}$ into luminosity in real disks are selective to certain frequencies (perhaps set by the local dissipation time-scale). This could potentially be the source of the frequency dependence observed in time lags measured from actual light curves. Going by this argument, $\dot{m}$ may not be a good proxy for luminosity when studying timing properties that involve dissipative processes. Propagating fluctuation models \citep{KCG01,AU06, MD18} get around this problem and recover frequency-dependent time lags by weighting the propagation time of the fluctuations by an emissivity profile that accounts for the conversion of mass accretion rate into luminosity. We will perform a similar procedure in Section 5.

\subsection{The rms-flux relationship}
\label{sec:rms-flux}
Observed light curves for several black hole binaries and AGNs display a linear relationship between the rms and the mean of the flux \citep{UM2001}, indicating that the brighter the source, the more variable it is. This linear rms-flux relation is considered to be a more important and fundamental characteristic of X-ray variability than power spectra for two main reasons: a) this relationship is observed during all spectral states of BHXRBs, while the PSD shape evolves during these spectral transitions \citep{GWPUNS2004}; and b) PSD alone cannot distinguish between different variability models, while the rms-flux relation can.

\begin{figure*}
\centering
\includegraphics[width=\textwidth]{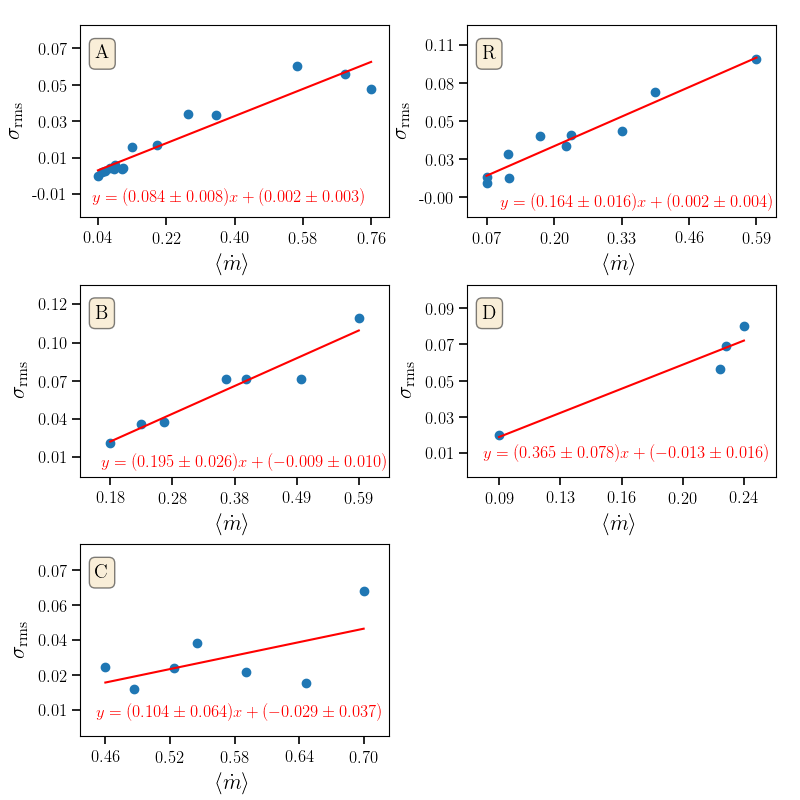}
\caption{rms-mean relation computed for $\dot{m}$ at the ISCO for all five simulations. Fits using equation~\ref{eqn:linearrms} are provided for each case.}
\label{fig:mdot_risco_rms}
\end{figure*}

Even a casual examination of Fig.~\ref{fig:mdot_rin} reveals that the amplitude of $\dot{m}$-variability in all our simulations increases and decreases in proportion to the magnitude of $\dot{m}$, consistent with an rms-mass flux relationship. To test this formally, we compute the rms-mean flux relationship for $\dot{m}$ at the ISCO, (which is $6\rg$ for all simulations except simulation D, which has its ISCO at $4.24\rg$). The usual way to compute the rms-flux relation is to divide the $\dot{m}$ data into a certain number of bins, and then compute the mean and absolute rms of $\dot{m}$ in each bin, which can then be plotted to get the relationship. This method obviously does not discriminate between the propagating fluctuations assumed in the model and possible non-dissipative, high-frequency fluctuations. One way we could filter out the non-dissipative fluctuations while computing the rms-mean relationship is to compute the rms using power spectra. Since we compute the root-mean square normalised power spectra, one can obtain the rms by simply integrating the power over the desired frequency range and taking the square root of it. Our radial coherence plots and time lag plots strongly suggest that below the local viscous frequency, dissipative fluctuations propagate inwards, which sets the upper limit on the relevant frequency range. Following a procedure similar to \citet{UM2001}, we divide the $\dot{m}$ data into a certain number of bins and for each bin we compute its periodogram, as described in section~\ref{sec:psd}, except that now we do not average the periodograms obtained over all the segments, but we do average the periodograms over each logarithmically-spaced frequency bin to obtain the (rms/mean$)^2$ power spectra. Next, we multiply the spectrum by the squared, mean value of $\dot{m}$ in that bin and integrate over all the frequencies below the viscous frequency. Finally, we take the square root of the result to recover the rms value. We repeat this procedure for each bin and then plot the results in Fig.~\ref{fig:mdot_risco_rms}. For all simulations except D, we use the same binning as for their radius-frequency power spectra, i.e. segments of length $0.13\,(M/M_{\odot})$~s for simulations A, B, and C, and $0.11\,(M/M_{\odot})$~s for simulation R. Since we need more than two points to look for a trend, we consider smaller segments of length $0.06\,(M/M_{\odot})$~s for simulation D. Data from each simulation are fit with the following linear function using a least-square method \citep{UM2001}:
\begin{equation}
    \sigma_{\rm rms} = k \langle\dot{m}\rangle+C.
    \label{eqn:linearrms}
\end{equation}
The resulting relations along with their least-square fits are reported in Fig.~\ref{fig:mdot_risco_rms}.

Simulations A, B, and R, seem to show strong linear rms-mass flux relations, as observed in nature. However the slopes are smaller than what is typically observed in BHXRBs. For Cygnus X-1, for example, the observed slope in the low/hard state is close to 0.3, and only decreases to around $\sim 0.16$ in the high/soft state~\citep{GWPUNS2004}. Simulation D exhibits the highest slope at $\approx 0.4$ (but perhaps, this is not reliable as it is anchored by a single low-frequency point), followed by simulations B and R with slopes at $\approx 0.2$. Simulations A and C appear to show linear rms-$\dot{m}$ relations, although their slopes are quite small. For simulation A, this could be the result of the secular decline in $\dot{m}$. For simulation C, this could possibly be due to the initial rise followed by a decline in $\dot{m}$. We note that upon using the corrected $\dot{m}_{\rm cor}$ curves, the rms-$\dot{m}$ relation is actually more scattered, possibly due to residuals between the actual decay and the exponential fit.

\subsection{Log-normal distribution}
\label{sec:log-normal}
Another signature of an underlying non-linear variability process is a log-normal distribution of the observed flux \citep{UMV2005}. The presence of a log-normal distribution indicates that the variability is caused by a multiplicative process, which here happens to be the coupling between fluctuations at different radii. Before testing our distributions of $\dot{m}$ for log-normality, it is important to note that the phenomenological models used to explain the log-normal distribution of flux assume that the light-curve is at least weakly stationary, i.e., the mean and variance do not change significantly within the time window of interest, while the higher ``moments'' may change. As we noted in earlier sections, $\dot{m}$ for most of our simulations undergoes a secular decline, implying it is not stationary over the period we wish to analyse. For this reason, we compute the distributions of $\dot{m}$ at the horizon only for data segments during which $\dot{m}$ is reasonably stationary\footnote{We select these segments by ``eye,'' so the results may not be representative of the statistical results of the full data.} as shown in Fig.~\ref{fig:hist}. In Appendix~\ref{appendix:b}, we show what the distributions look like if the entire simulation duration is considered. 

In Fig.~\ref{fig:hist}, in the top row of each panel, we show the $\dot{m}$ segments used for each simulation. Below that, we show the resulting histograms, fitted with both normal (red, dashed curve) and log-normal (blue, solid curve) distributions. The log-normal fit is made using
\begin{equation}
    f(x;\mu,\sigma,\tau) = \dfrac{1}{\sqrt{2 \pi}\sigma (x-\tau)}\exp \left\{-\dfrac{\left[\log(x-\tau)-\mu\right]^2}{2\sigma^2}\right\},
    \label{eq:log-normal}
\end{equation}
with $\tau$ as an offset parameter. Best-fit parameters for these distributions are provided in Table~\ref{tab:mdot_rin_hist}.

\begin{table}
\centering
\begin{tabular}{m{0.25cm} m{0.35cm} m{0.35cm} m{1cm} m{0.65cm} m{0.32cm} m{0.65cm} m{1cm}}
\hline
\multicolumn{1}{c}{}  & \multicolumn{3}{c}{Normal} & \multicolumn{4}{c}{Log-normal} \T\B \\\hline
\multicolumn{1}{c}{Sim} & \multicolumn{1}{c}{$\mu$} & \multicolumn{1}{c}{$\sigma$} & \multicolumn{1}{c}{$\chi^2/$d.o.f.} & \multicolumn{1}{c}{$\mu$} & \multicolumn{1}{c}{$\sigma$} & \multicolumn{1}{c}{$\tau$} & \multicolumn{1}{c}{$\chi^2/$d.o.f.}\T\B\\\hline
\multirow{3}{*}{A} & 0.57 & 0.06 & 252.4/13 & -2.44 & 0.58 & 0.47 & {23.8/12} \T\B \\ \cline{2-8} 
& 0.09 & 0.005& 182.5/14& -4.0 & 0.28& 0.07 & {14.7/13} \T\B\\ \cline{2-8} 
& 0.03& 0.002 & 609.4/17 & -5.0 & 0.30 & 0.03 & {13.8/16} \T\B\\\hline
\multirow{3}{*}{B} & 0.46 & 0.13 & 277.8/16 & -1.25 & 0.41 & 0.15 & {16.0/15} \T\B\\ \cline{2-8} 
& 0.39 & 0.08& 71.8/16& -1.16 & 0.24& 0.07 &{28.1/15} \T\B\\ \cline{2-8} 
& 0.2& 0.05 & 86.1/16 & -2.11 & 0.38 & 0.07 & {32.1/15} \T\B\\\hline
\multirow{1}{*}{C}& 0.49& 0.02& 11./13 & -0.38 & 0.03 & -0.2 & 11.1/12 \T\B\\ \hline
\multirow{3}{*}{R} &0.18& 0.04& 1459.1/38 & -2.29 & 0.36 & 0.07 & \multicolumn{1}{c}{217.0/37}\T\B\\  \cline{2-8} 
&0.11 & 0.02 & 1257.4/42 & -2.1& 0.2 & -0.02 & \multicolumn{1}{c}{278.2/41}\T\B\\  \cline{2-8} 
&0.05& 0.02& 1726.1/42 & -3.38 & 0.42 & 0.01& \multicolumn{1}{c}{216.8/41}\T\B\\\hline
\end{tabular}
\caption{Parameters for the normal and log-normal fits to the mass flux histograms, corresponding to the plots in Fig.~\ref{fig:hist}.}
\label{tab:mdot_rin_hist}
\end{table}

Although simulation A shows significant decline in $\dot{m}$, there are sufficiently long time segments during which $\dot{m}$ is stationary enough to reasonably expect a well-defined distribution. We found that, indeed, the $\dot{m}$ distributions in these segments are statistically well fit by a log-normal distribution, with $\chi^2/\mathrm{d.o.f.} = 1.98$, 1.13 and 0.86. This is not true for the entire simulation data (see Fig.~\ref{fig:mdot_rin_hist_all}). 

Similar to simulation A, the shorter time segments of simulation B show statistically significant log-normal distributions, with $\chi^2/\mathrm{d.o.f.} = 1.10$, 1.87 and 2.14. Even though, in this case, the secular decline is not very steep, the entire simulation data are still not well fit by a log-normal distribution (see Table~\ref{tab:mdot_rin_hist_all}). None of the data segments give an acceptable fit for a Gaussian distribution (see Table~\ref{tab:mdot_rin_hist}).  

In the case of simulation C, the steady increase in $\dot{m}$ initially, followed by its gradual decline makes it hard to find a reasonably long time segment in which  $\dot{m}$ is stationary. In Fig.~\ref{fig:hist}, we show a segment chosen towards the end of the simulation, where $\dot{m}$ is nearly stationary. In contrast to the previous results, we find this distribution to be well fit by either a normal or log-normal distribution. 

For different time segments in simulation R, we find that both visibly and based on $\Delta \chi^2$ values, a log-normal distribution fits better than a normal one. However, the obtained $\chi^2/$d.o.f. for the log-normal fits are not statistically acceptable. A similar conclusion is reached when the entire duration is considered (see Table~\ref{tab:mdot_rin_hist_all}).

Simulation D does not exhibit as significant a decline as the other simulations, so we show its entire distribution of $\dot{m}$ in Fig.~\ref{fig:mdot_rin_hist_all}, along with the other simulations. This simulation is well-fit by a log-normal distribution with $\chi^2/\mathrm{d.o.f.} = 1.12$.

\begin{figure*}
\begin{subfigure}[t]{1.0\textwidth}
\centering
\includegraphics[scale=0.5]{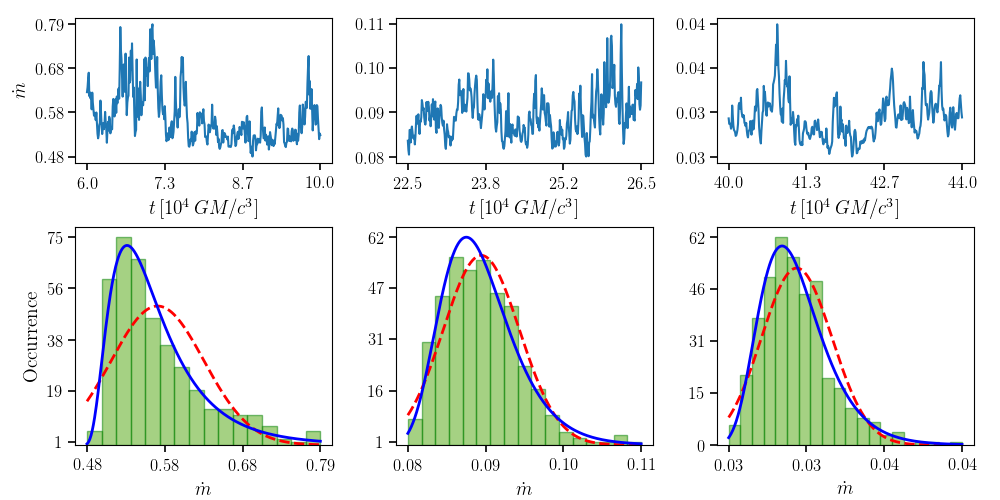}
\label{fig:histA}
\end{subfigure}
\\
\begin{subfigure}[t]{0.74\textwidth}
\includegraphics[scale=0.51]{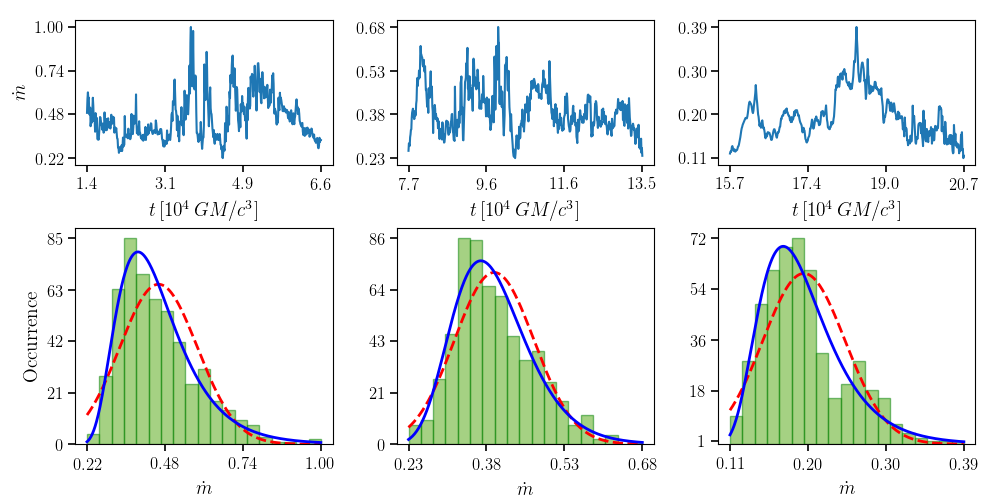}
\label{fig:histB}
\end{subfigure}
\begin{subfigure}[t]{0.24\textwidth}
\includegraphics[scale=0.51]{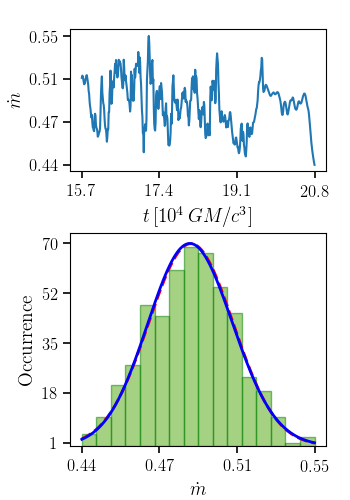}
\label{fig:histC}
\end{subfigure}
\\
\begin{subfigure}[t]{1.0\textwidth}
\centering
\includegraphics[scale=0.5]{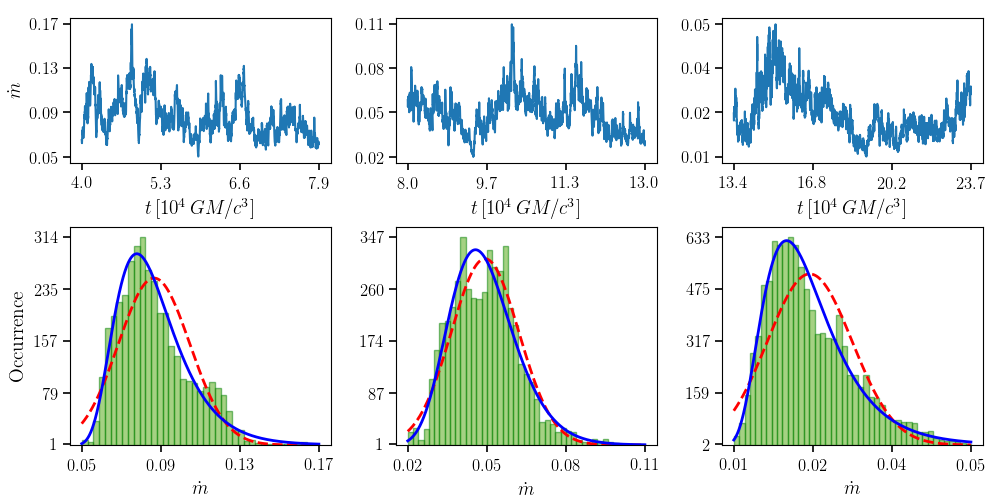}
\label{fig:histR}
\end{subfigure}
\caption{Distribution of $\dot{m}$ for simulations A, B, C and R for different time segments during which $\dot{m}$ is reasonably stationary. The best fits of normal and log-normal distributions are shown in red, dashed and blue, solid curves, respectively.}
\label{fig:hist}
\end{figure*}

\section{Comparison with observations}
\label{sec:obs}

One must take care when comparing the results of these simulations to light curves, as of course observed light curves are not a result of just the mass accretion rate profiles present in the flow, but a product of the $\dot{m}$ time series with their associated spectral emission components. Indeed, the mass accretion rate behaviour at a given radius may be effectively invisible to observers if there is no corresponding emission mechanism at that location, or the variability at a given radius may be greatly amplified in observations if the relevant emission mechanism at that radius happens to dominate the observed energy band. Drawing direct analogies to observed light curves is, therefore, risky, but we can begin to draw some qualitative comparisons nevertheless.

The power spectral profile exhibited by simulations A and R, in particular (Figs. \ref{fig:psdradii_A} and \ref{fig:psdR_radii}), show many interesting points of overlap with observation. As in these simulations, an abundance of power over a broad frequency range with a break at high-frequencies is commonly observed in X-ray binary light curves at low accretion rates \citep[e.g.,][]{GZD97, Axelsson2008}. This high-frequency break has generally been attributed to the viscous frequency at the inner edge of the accretion flow, although interference effects may lower the actual frequency \citep{ID2011}. In this picture, it would be strange then for the power from 2, 10 and $20\rg$ to have very similar break frequencies. However we note that observations of light curves in different energy bands from a given source \citep{WU09, DMPN2015} exhibit high frequencies breaks in the same range, often with no apparent correlation with band centroid energy consistent with our simulation results. 

\begin{figure}
\centering
\includegraphics[scale=0.52]{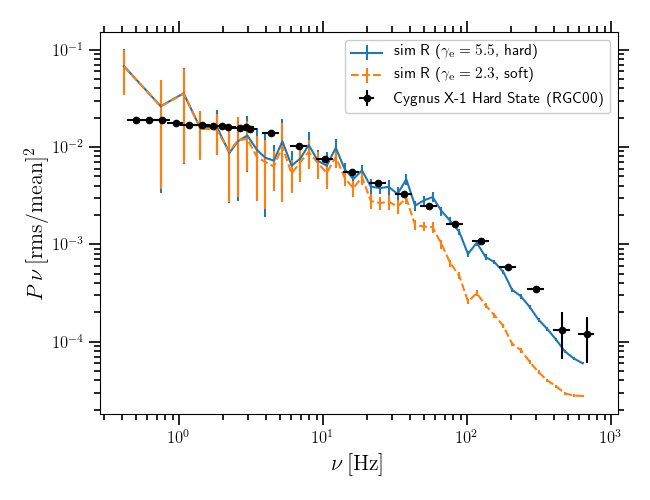}
\caption{Power spectra of simulation R for hard and soft energy bands, overlayed with the observational data of Cygnus X-1 in the hard state, taken from \citet{RGC2000} (RXTE ObsID P30157). Here, we assume $M_\mathrm{Cyg~X-1} = 15\,M_\odot$.}
\label{fig:RCygX1}
\end{figure}

\begin{figure}
\centering
\includegraphics[scale=0.52]{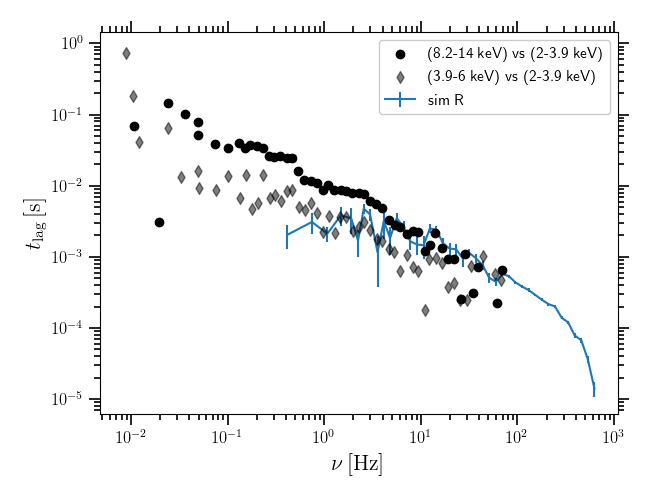}
\caption{Lag spectrum of hard band vs soft band for simulation~R, compared with the observed lags in low/hard state of Cygnus X-1 \citep{NVBW99, P2001}. Here, we assume $M_\mathrm{Cyg~X-1} = 15\,M_\odot$.}
\label{fig:RCygX1_lag}
\end{figure}
The presence of inter-annular time lags also represents an excellent bridge between simulations and observations. The model for how inter-band time lags arise in observed accretion flows requires that fluctuations generated in the outer disc first liberate energy there, before propagating inward to modulate emission at smaller radii and produce the observed lag \citep{L97}. For a typical $10$~$M_{\odot}$ BH, Fig.~\ref{fig:time_lags} and Table~\ref{tab:tlags} indicates lags of the order of 1-10~ms, which are comparable to the lags observed between low and high energy X-ray bands in many sources in the hard state \citep{DPPC17, Mahmoud18b}.

One aspect missing from the GRMHD results is the power-law decay in time lag with frequency commonly seen in observations. In our simulations, lags are simply explained by the propagation time from one radius to another and therefore are frequency-independent. However, the dissipative processes that convert $\dot{m}$ into luminosity could potentially introduce some frequency-dependence in the lags. Often in models, these dissipative processes are well captured by radially dependent emissivity profiles \citep[see for example][]{ID2011, IK2013}. Here we follow a similar procedure and compute the emissivity-weighted flux observed in a particular energy band as 
\begin{equation}
    f(t) = \sum_{r =r_{\rm H}}^{r_{\rm out}} \epsilon (r)\,\dot{m}(r,t)\, ,
    \label{eq:flux}
\end{equation}
where we take the emissivity profile $\epsilon (r) = ({\rm d}r/r)\,r^{2-\gamma_{\rm e}}$, $r_{\rm out} =25\rg$, and $\gamma_{\rm e}$ is the emissivity index, which is a free parameter. The higher-energy band is expected to have a steeper emissivity profile compared to the soft-energy band; therefore, we take $\gamma_{\rm e}$ = 5.5 and 2.3 for the hard and soft energy bands, respectively. Fig.~\ref{fig:RCygX1} shows the power spectra of the thus-obtained light curves in the hard and soft energy bands for simulation R. Within the measurable range, the simulation hard band power spectrum is consistent with the power seen in, e.g., the hard states of Cygnus X-1 \citep{RGC2000}. Note that, due to the limited duration of these simulations, our power spectra can only go down to $\nu\sim 10\,(M_{\odot}/M)$~Hz. This means we cannot probe the range where one typically sees what is called the low-frequency break, $\nu\sim 0.1$ to $1\,(M_{\odot}/M)$~Hz.

We can also use our synthetic light curves to measure the time lags between the hard and soft bands. Results for simulation R are shown in Fig.~\ref{fig:RCygX1_lag}. Unlike the lags measured from coherent $\dot{m}$ fluctuations between any two radii, which are frequency independent (see Fig.~\ref{fig:time_lags}), these lags vary with the frequency (e.g., declining monotonically, within error, between 5 and 100 Hz in Fig.~\ref{fig:RCygX1_lag}) and match well with the lag spectra between different energy bands for Cygnus X-1 in the hard-spectral state \citep{NVBW99, P2001}. This simple exercise indicates that dissipative processes could play a vital role in modulating the underlying variability in $\dot{m}$, as it is imprinted on the light-curve. GRMHD simulations that include radiative processes will be needed to explore this idea further.

Taken together, the presence of inter-annular time lags of magnitude matching the observations, the presence of high frequency breaks in many of the power spectra, and the systematic trends in the coherence represent an excellent starting point from which observers may begin to interpret real light curves in terms of the underlying $\dot{m}$ behaviour.

\section{Conclusions}
\label{sec:conclusions}
In this work, we tested the hypothesis that the broadband X-ray variability observed in black hole X-ray binaries and AGN is attributable to mass-flux variations in the accretion flow. We did this through analysis of a suite of long-duration, GRMHD simulations. We considered the mass accretion rate as a proxy for luminosity to identify multiple variability features. We found that, despite the differences in the initial set ups and outcomes of the simulations, each one showed evidence for inward propagating fluctuations.
Our findings include:
\begin{itemize}
\item In general, the power spectra from these simulations have two components: low frequency power below the viscous frequency and high frequency power above the Keplerian frequency, with the exception of simulation C, which shows relatively less $\dot{m}$ variability.

\item Simulations show evidence for power along and above the Keplerian frequency curve. The presence of this power causes the high frequency power at larger radii to dominate that at smaller radii. While this is in contrast to X-ray observations, in which higher energy bands exhibit more high frequency power compared to lower energy bands, the presence of the high frequencies only above the Keplerian curve strongly suggests they are associated with $p$-modes, which are not likely to strongly modulate the light curve. Perhaps, with proper treatment of radiative processes (i.e., radiative GRMHD simulations), we would be able to filter out the non-dissipative, high-frequency fluctuations by obtaining the power spectra directly from the luminosity.

\item The $\dot{m}$ power spectra at a given radius exhibit strong similarity with spectra at smaller radii below the local viscous frequency. This is in contrast to model assumptions that emission at each radius peaks at the local viscous frequency. 

\item All simulations show remarkable radial coherence below the viscous frequency. 
This agrees well with how most models explain the strong coherence between energy bands.

\item All simulations display positive time lags below the viscous frequency, when comparing the fluctuations of $\dot{m}$ at smaller radii with those at larger radii. This, together with the radial coherence, strongly supports the presence of inward propagating fluctuations in the accretion flow.  

\item The time lags between highly coherent fluctuations of $\dot{m}$ between any two fixed radii are frequency {\em independent}.
\item When they are measured between two synthetic energy bands, generated using power-law emissivity profiles, the same $\dot{m}$ fluctuations yield frequency {\em dependent} time lags, consistent with X-ray observations.

\item All simulations show a linear rms-mean $\dot{m}$ relation, although the slopes obtained are smaller than typically observed from linear rms-mean flux relations in BHXRBs.

\item All simulations, except R, show log-normal behaviour within those time segments in which $\dot{m}$ is approximately stationary. This strongly supports the notion that the fluctuations arise from a multiplicative, stochastic process. 

\end{itemize}

In conclusion, this work is a promising step toward confirming the propagating fluctuation model and connecting simulations with observations.

\section*{Acknowledgements}
We thank the anonymous referee for the helpful comments on the paper. We acknowledge Alex Markowitz, Barbara De Marco, and Omer Blaes for helpful discussions. This research was partly supported by the Polish NCN grants 2018/29/N/STP/02134, UMO-2018/29/N/ST9/02139 and 2019/33/B/ST9/01564 and Polish National Agency for Academic Exchange grant PPN/IWA/2018/1/00099/U/0001. Computations for simulations A, B and C used the Extreme Science and Engineering Discovery Environment (XSEDE) Stampede2 at the Texas Advanced Computing Center through allocation AST170012, as well as the Savio computational cluster resource provided by the Berkeley Research Computing program at the University of California, Berkeley. Computations for simulation~D were performed on the Prometheus cluster, part of the PL-Grid infrastructure located at ACK Cyfronet AGH in Poland. PCF appreciates the support of National Science Foundation grants AST1616185, PHY-1748958, and AST-1907850.

\appendix
\section{Details of simulation D}
\label{appendix:a}
This simulation is performed using a spherical-polar grid with the computational domain extending from $1.8$ to $80^{1.5}\rg$ in the radial direction. We set small cut-outs near the poles with $\theta \in [0.02\pi, 0.98\pi]$ to keep the timestep reasonable. In the azimuthal direction, we simulate only a $\pi/2$ wedge. We use logarithmic spacing in the radial direction of the form
\begin{equation}
    x_1 = 1 +\ln\left(\frac{r}{r_{\rm H}}\right) ~.
\end{equation}
In the polar direction, we chose $\theta$ of the form \citep{Mckinney06}
\begin{equation}
    \theta = x_2+\dfrac{1}{2}[1-q] \sin(2 x_2) ~,
\end{equation}
where the parameter $q$ (set to 0.5) determines the concentration of grid zones near the equator. We set outflow boundary conditions (copying scalar fields to ghost zones, while ensuring the velocity component normal to the boundary points outward) at the inner and outer edges of the domain in both the radial and poloidal directions. We use periodic boundary conditions in the $\phi$-direction. 

We set the poloidal magnetic field by using a purely azimuthal vector potential of the form 
\begin{equation}
    A_{\phi}= S\left( \frac{\rho}{u^t \sqrt{-g}} -0.2\rho_{\rm max}\right)^2 \sin \left[ 2\log \left( \frac{r}{1.1r_{\rm in}}\right)\right] ~,
\end{equation}
where $S=1$ if $\theta<\pi$ and $S=-1$ if $\theta >\pi$; $\rho_{\max}$ is the maximum density of the torus, located at $r_{\rm max}$; $r_{\rm in}$ is the inner edge of the torus; and $u^t$ is the time component of the four-velocity. 

The Cosmos++ code uses an explicit five-stage, strong-stability-preserving Runge-Kutta time integration scheme \citep{SR02}, set to third order. Conserved variables are updated at each time step using the high resolution shock-capturing (HRSC) method. The reconstruction of the primitive variables at different spatial locations is done through Piecewise Parabolic Method (PPM) interpolation with a monotonized central limiter \citep{CW84}. This is followed by the calculation of flux terms at the zone faces using the Harten-Lax-van Leer (HLL) Riemann solver. In order to maintain a divergence-free magnetic field during the evolution, we use the staggered, constrained transport scheme described in \citet{FGMRA12}. For the primitive inversion step, we primarily use the ``2D'' solver from \citet{NGM06}, with a 5D numerical inversion scheme as a backup. In cases where both solvers fail, we use the entropy instead of the conserved energy to recover the primitive variables.

\begin{figure}
\centering
\includegraphics[scale=0.6]{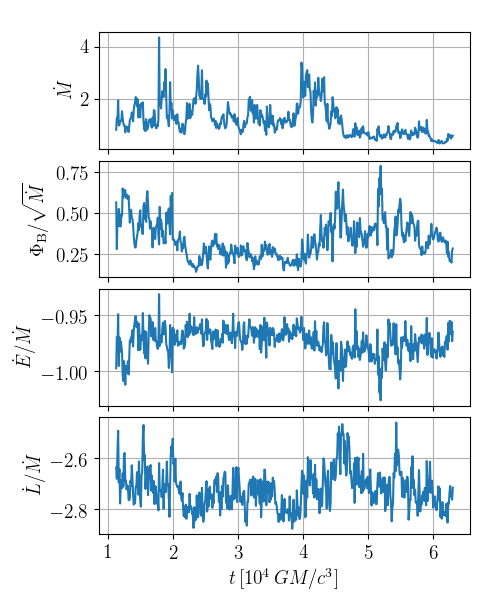}
\caption{Evolution of mass accretion rate (arbitrary units; top panel), magnetic flux (second panel), energy flux (third panel) and angular momentum flux (bottom panel) through the horizon for simulation D.}
\label{fig:D_fluxes}
\end{figure}

Since this simulation has not been described elsewhere in the literature, we report some basic diagnostics here. First, in Fig.~\ref{fig:D_fluxes}, we plot various fluxes through the event horizon. We define the magnetic flux ($\Phi_{\rm B}$), energy flux ($\dot{E}$) and angular momentum flux ($\dot{L}$) as:
\begin{eqnarray}
\Phi_{\rm B} &=& \dfrac{1}{2}\oint |B^r| \sqrt{-g} d\theta d\phi ~,\\
\dot{E} &=& -\oint T^r_{t} \sqrt{-g} d\theta d\phi ~,\\
\dot{L} &=& \oint T^r_{\phi} \sqrt{-g} d\theta d\phi ~,
\label{eq:phib}
\end{eqnarray}
where $B^r$ is the radial-component of the primitive magnetic field, which is already scaled by the $\sqrt{4\pi}$ factor. $T^r_t$ and $T^r_{\phi}$ are components of the stress energy tensor \citep[see eq. 4 in][]{FGMRA12}. The modest value of the magnetization parameter, $\Phi_{\rm B}/\sqrt{\dot{M}}$, throughout the simulation indicates that we maintain a SANE (standard and normal evolution) flow, even though we initiate the simulation with a single poloidal field loop. These fluxes generally agree well with the other simulations, especially R \citep{NSPK2012}. The one exception is that the magnetic flux through the horizon is somewhat lower than the others, likely due to the choice of boundary conditions in Cosmos++ \citep[see, e.g.,][]{EHT2019}. 

\begin{figure}
\centering
\includegraphics[scale=0.27]{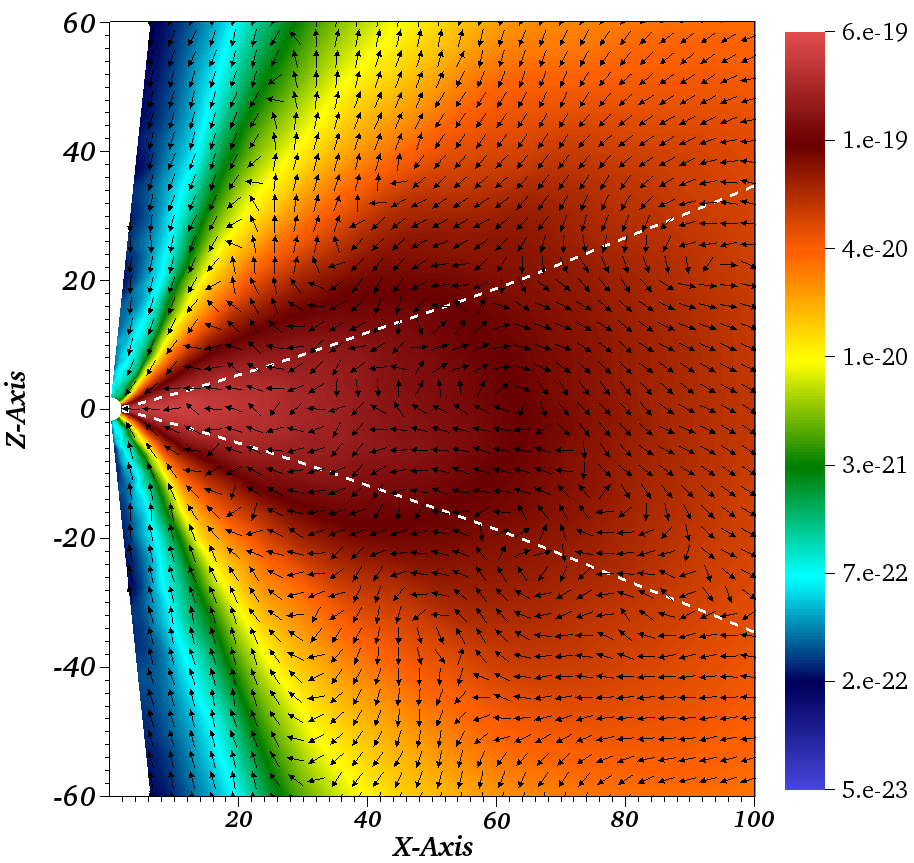}
\caption{Azimuthally- and time-averaged velocity vectors with the background color representing the similarly averaged density. The white, dashed line indicates the time-average of the density scale height. Time averaging is done over the last $10,000\,GM/c^3$ of the simulation.}
\label{fig:D_tavg_rho}
\end{figure}

In Figure~\ref{fig:D_tavg_rho}, we show the azimuthally- and time-averaged, velocity vectors superimposed on the averaged density. Time averaging is performed for the last $10,000\,GM/c^3$ period of the simulation. The white, dashed line represents the time average of the density scale height. The resulting accretion flow is clearly geometrically thick, with dominant inflow of matter at all latitudes within $\sim 20\rg$ unlike simulations A, B and C which exhibit outflows/convective motions. Beyond this radius, there are signatures of outflows at higher latitudes. Small convective loops can be noticed in some locations. In general, these results are similar to simulation~R. 
 
\section{Log-normal distribution for full $\dot{m}$}
\label{appendix:b}
\begin{figure*}
\centering
\includegraphics[width=\textwidth]{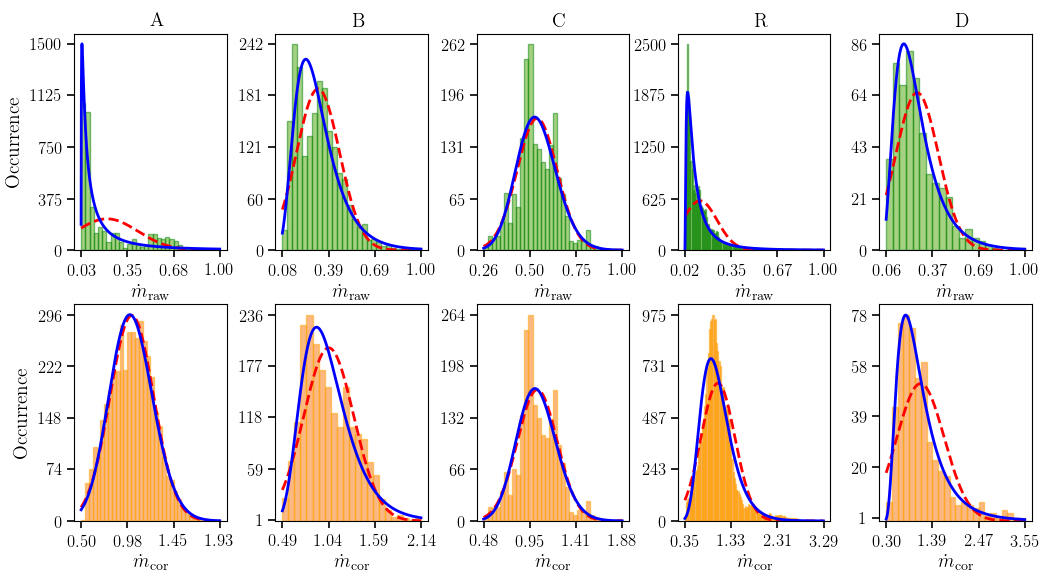}
\caption{Histograms of the normalised accretion rate ($\dot{m}$) at the horizon ($r_{\rm H}$) for all five simulations. Top row: Histograms for the un-corrected $\dot{m}$. Bottom row: Same for the corrected accretion rates, $\dot{m}_\mathrm{cor}$. Best-fits normal and lognormal distributions are shown in red, dashed and blue, solid curves, respectively.}
\label{fig:mdot_rin_hist_all}
\end{figure*}

In Section \ref{sec:log-normal}, we analysed histograms of $\dot{m}$ selected for periods when the mean accretion rate was reasonably constant. Here we repeat the analysis for the full distribution of $\dot{m}$ for every simulation, excluding the discarded initial transient data (see Fig.~\ref{fig:mdot_rin}). Fig.~\ref{fig:mdot_rin_hist_all} shows the resulting histograms. In the top row, we show the distribution for the original $\dot{m}$ data, with the best fits for normal and log-normal distributions shown in red, dashed and blue, solid curves, respectively. We find that, except for simulation~D, none of the fits have a statistically significant $\chi^2/$d.o.f. value (see Table \ref{tab:mdot_rin_hist_all}). To test whether subtracting off a simple exponential fit to the raw $\dot{m}$ could improve the fit, we also present histograms of $\dot{m}_\mathrm{cor}$ in the bottom panel of Fig.~\ref{fig:mdot_rin_hist_all}. Best-fit parameters for all the distributions are provided in Table~\ref{tab:mdot_rin_hist_all}, along with the respective $\chi^2/$d.o.f. values for the fits. Although the fits for $\dot{m}_\mathrm{cor}$ do show smaller $\chi^2/$d.o.f. values, they are still not statistically acceptable.  
We re-emphasize, though, that Fig. \ref{fig:hist} and Table \ref{tab:mdot_rin_hist} show that whenever $\dot{m}$ oscillates around a time-steady mean, the fluctuations {\em are} well fit by a log-normal distribution.

\begin{table*}
\centering
\caption{Parameters for the normal and log-normal fits to the mass flux histograms in Fig.~\ref{fig:mdot_rin_hist_all}}
\begin{tabular*}{1 \textwidth}{ccccccccc}
\cline{1-9}
\multicolumn{2}{c}{Sim}  & \multicolumn{3}{c}{Normal} & \multicolumn{4}{c}{log-normal}\T\B\\ \cline{1-9}
& &  $\mu$  & $\sigma$ & $\chi^2/$d.o.f. & $\mu$   & $\sigma$ & $\tau$  & $\chi^2/$d.o.f.\\ \cline{3-9}
\multirow{2}{*}{A} &  \multicolumn{1}{c}{$\dot{m}$}&  0.21 & 0.22 & 9880.4/30 & -2.64& 1.57& 0.03 &2067.1/29 \T\B\\\cline{2-9}
 &\multicolumn{1}{c}{$\dot{m}_{\rm cor}$} &1.03& 0.22&117.8/33& 1.32&0.06&-2.72&127.3/32\T\B\\\cline{1-9}
\multirow{2}{*}{B} &  \multicolumn{1}{c}{$\dot{m}$}& 0.32 &0.14 & 1632.3/25&-1.19&0.43&-0.01&216.2/24\T\B\\\cline{2-9}
&\multicolumn{1}{c}{$\dot{m}_{\rm cor}$} &1.04&0.30&281.5/20&-0.22&0.35&0.19&150.3/19\T\B\\\cline{1-9}
\multirow{2}{*}{C}& \multicolumn{1}{c}{$\dot{m}$} &0.54& 0.11 & 509.1/31 & 0.24 & 0.08 & -0.74 &368.1/30\T\B\\\cline{2-9} 
&\multicolumn{1}{c}{$\dot{m}_{\rm cor}$}& 1.02 & 0.20 &491.4/31 & 0.93 & 0.08 & -1.51 & 351.7/30\T\B\\\cline{1-9}
\multirow{2}{*}{R} &  \multicolumn{1}{c}{$\dot{m}$} & 0.13 & 0.12 & 532778977.8/122 & -2.77 & 1.07 & 0.02 & 1288.5/121\T\B\\\cline{2-9} 
&\multicolumn{1}{c}{$\dot{m}_{\rm cor}$} & 1.05 & 0.36 & 729349.7/113& 0.02 & 0.31 & -0.02 & 6454.4/112\T\B\\ \cline{1-9}
\multirow{2}{*}{D}&\multicolumn{1}{c}{$\dot{m}$} & 0.27 & 0.14 & 2353.8/18 & -1.4& 0.52 & -0.01& 19.1/17\T\B\\\cline{2-9} 
&\multicolumn{1}{c}{$\dot{m}_{\rm cor}$}& 1.11 & 0.55 & 792.8/21 & -0.27 & 0.57 & 0.21& 25.8/20\T\B\\\cline{1-9}
\end{tabular*}
\label{tab:mdot_rin_hist_all}
\end{table*}

\section*{Data Availability}
The data underlying this article will be shared on a reasonable request to the corresponding author.
\bibliographystyle{mnras}
\bibliography{variability} 
\bsp	
\label{lastpage}
\end{document}